\documentclass[a4paper,10pt,twoside,titlepage]{book}
\usepackage[dvips]{epsfig}
\usepackage{amsmath}
\usepackage{amssymb} 

\newcommand{\be}{\begin{equation}}\newcommand{\ee}{\end{equation}}\newcommand{\ba}{\begin{eqnarray}}\newcommand{\ea}{\end{eqnarray}}\newcommand{\ban}{\begin{eqnarray*}}\newcommand{\ean}{\end{eqnarray*}}

\newcommand{\braket}[2]{\mbox{$ \langle #1 | #2 \rangle $}}
\newcommand{\sandwich}[3]{\mbox{$ \langle #1 | #2 | #3 \rangle $}}
\newcommand{\ket}[1]{\mbox{$ | #1 \rangle $}}
\newcommand{\bra}[1]{\mbox{$ \langle #1 | $}}
\newcommand{\com}[2]{\left[ #1\,,\,#2 \right]}

\newcommand{\si}{\sigma}

\newcommand{\demi}{\frac{1}{2}}
\newcommand{\compl}{\begin{picture}(8,8)\put(0,0){C}\put(3,0.3){\line(0,1){7}}\end{picture}}
\newcommand{\real}{\begin{picture}(8,8)\put(0,0){R}\put(0,0){\line(0,1){7}}\end{picture}}

\newcommand{\one}{\leavevmode\hbox{\small1\normalsize\kern-.33em1}}%

\newcommand{\Tr}{\mbox{Tr}}

\newcommand{\moy}[1]{\langle #1 \rangle}

\begin{document}

\frontmatter{\title{Les Houches school of Physics in Singapore\\ \Huge{\bf{QUANTUM INFORMATION}}\\ \Large{\bf{Primitive notions and quantum correlations}}}

\author{{\bf Valerio Scarani}\\Centre for Quantum Technologies and Department of Physics\\
National University of Singapore\\physv@nus.edu.sg}}
\maketitle

\tableofcontents%
\mainmatter

\chapter{Quantum theory}

This Lecture covers basic material, that can be found in many sources. If I'd have to suggest some general bibliography, here are my preferences of the moment: for quantum physics in general, the book by Sakurai \cite{sakurai}; for a selection of important topics that does not follow the usual treatises, the book by Peres \cite{peres}; and for quantum information, the books of Nielsen and Chuang \cite{nielsen} or the short treatise by Le Bellac \cite{lebellac2}.

\section{Axioms, assumption and theorems}

\subsection{Classical versus quantum physics}

It is convenient to start this series of lectures by reminding the meaning of some frequently used expressions:

\textit{Physical system:} the degrees of freedom under study. For instance, the system ``point-like particle'' is defined by the degrees of freedom $(\vec{x},\vec{p})$. A system is called \textit{composite} if several degrees of freedom can be identified: for instance, ``Earth and Moon''.\\
\textit{State at a given time:} the collection of all the properties of the system; a \textit{pure state} is a state of maximal knowledge, while a \textit{mixed state} is a state of partial knowledge.

An extremely important distinction has to be made now: the distinction between the \textit{description} (or kinematics) of a physical system and its \textit{evolution} (or dynamics). The distinction is crucial because \textit{the specificity of quantum physics lies in the way physical systems are described, not in the way they evolve!}

\subsubsection{Classical physics}

In classical physics, the issue of the description of a physical system is not given a strong emphasis because it is, in a sense, trivial. Of course, even in classical physics, before studying ``how it evolves'', we have to specify ``what it is'', i.e. to identify the degrees of freedom under study. But once the system is identified, its properties combine according to the usual rules of set theory. Indeed, let $P$ represent properties and $s$ pure states. In classical physics, $s$ is a point in the configuration space and $P$ a subset of the same space. Classical logic translates into the usual rules of set theory:
\begin{itemize}
\item The fact that system in state $s$ possesses property $P$ translates as $s\in P$;
\item For any two properties $P_1$ and $P_2$, one can construct the set $P_1\cap P_2$ of the states that possess both properties with certainty;
\item If a property $P_1$ implies another property $P_2$, then $P_2\subset P_1$;
\end{itemize}
and so on. In particular, by construction $s=\bigcap_kP_k$ where $P_k$ are all the properties that the system possesses at that time. As well-known, only part of this structure can be found in quantum physics: in particular, a pure state is specified by giving the list of all \textit{compatible} physical properties that it possesses with certainty.

Another interesting remark can be made about composite systems. Let us consider for instance ``Earth and Moon'': the two subsystems are obviously interacting, therefore the dynamics of the Earth will be influenced by the Moon. However, at each time one can consider the state of the Earth, i.e. give its position and its momentum. If the initial state was assumed to be pure, the state of each sub-system at any given time will remain pure (i.e., the values sharply defined). This is not the case in quantum physics: when two sub-systems interact, generically they become ``entangled'' in a way that only the global state remains pure.

\subsubsection{Quantum physics}
 
In all these lectures, I shall focus on \textit{non-relativistic quantum physics} and use the following definition of quantum physics: 
\begin{itemize}
\item[(i)] Physical systems (degrees of freedom) are described by a Hilbert space;
\item[(ii)] The dynamics of a closed system is reversible.
\end{itemize}
Admittedly, (i) is not a very physical definition: it is rather a description of the formalism. Several attempts have been made to found quantum physics on more ``physical'' or ``axiomatic'' grounds, with some success; but I consider none of them conclusive enough to be worth while adopting in a school. We shall come back to this issue in Lecture 5. Requirement (ii) is not specific to quantum physics: in all of modern physics, irreversibility is practical but not fundamental (the result of the system interacting with a large number of degrees of freedom, which one does not manage to keep track of).

\subsection{Kinematics: Hilbert space}

\textit{Note:} the students of this school are supposed to be already familiar with the basic linear algebra used in quantum physics, as well as with the Dirac notation. Therefore, I skip the mathematical definitions and focus only on the correspondence between physical notions and mathematical objects.

\subsubsection{Pure states, Born's rule}

Basically the only statement that should be taken as an axiom is the description of states in quantum physics: the space of states is not a set, like the classical configuration space, but a \textit{vector space with scalar product}, defined on the complex field, called \textit{Hilbert space}\footnote{Strictly speaking, the Hilbert space is the infinite-dimensional vector space used in quantum physics to describe the point-like particle. But it has been customary to extend the name to all the vector spaces of quantum theory, including the finite-dimensional ones on which we mostly focus here.} and written ${\mathcal H}$. In this space, \textit{pure states are described by one-dimensional subspaces}. A one-dimensional subspace is identified by the corresponding projector $P$; or alternatively, one can choose any vector $\ket{v}$ in the subspace as representative, with the convention however that every vector $c\ket{v}$ differing by a constant represents the same state. 

With this construction, as well known, after a measurement a system may be found to possess a property that it did not possess with certainty before the measurement. In other words, given a state $\ket{v_1}$, there is a non-zero probability that a measurement finds the system in a \textit{different} state $\ket{v_2}$. The probability is given by
\ba
P(v_2|v_1)&=&\Tr\left(P_1\,P_2\right)\,=\,\left|\braket{v_1}{v_2}\right|^2
\ea
where we assume, as we shall always do from now onwards, that the vectors are normalized as $\braket{v}{v}=1$. This is called \textit{Born's rule for probabilities}.

What we have discussed in this paragraph is the essence of quantum physics. It is generally called \textit{superposition principle}: states are treated as vectors, i.e. the vector sum of states defines another state that is not unrelated to its components\footnote{The classical space of states has the structure of a set, not of a vector space, in spite of being identified with $\real^n$. Indeed, the vector sum is not defined in the configuration space: one can formally describe a point as the vector sum of two others, but there is no relation between the physical states (the ``sum'' of a car going East at 80km/h and a car going West at 80km/h is a parked car located at the mid-point). Yet another example of a set of numbers, on which operations are not defined, is the set of telephone numbers.}. All the usual rules of quantum kinematics follow from the vector space assumption. It is useful to sketch this derivation.

\subsubsection{Ideal measurements}

Consider a measurement, in which different positions of the pointer can be associated with $d$ different states $\ket{\phi_1},...,\ket{\phi_d}$. Among the features of an ideal measurement, one tends to request that the device correctly identifies the state. In quantum physics, this cannot be enforced for all states because of the Born's rule; but at least, one can request the following: if the input system is in state $\ket{\phi_j}$, the measurement should produce the outcome associated to that state with certainty (for non-destructive measurements, this implies that the measurement outcome is reproducible). Therefore, given Born's rule, \textit{an ideal measurement is defined by an orthonormal basis, i.e. a set of orthogonal vectors}.

A very compact way of defining an ideal measurement is provided by \textit{self-adjoint operators}, because any self-ajoint operator $A$ can be diagonalized on an orthonormal basis: $A=\sum_k a_k\ket{\phi_k}\bra{\phi_k}$, with $a_k$ real numbers. In this case, one typically considers $A$ to be the ``physical quantity'' that is measured and the $a_k$ to be the ``outcomes'' of the measurement. In turn, one is able to compute derived quantities, in particular average values over repeated measurements. Indeed, let $\ket{\psi}$ be the state produced by the source, $a^{(i)}$ be the outcome registered in the $i$-th measurement and $n_k$ the number of times the outcome $a_k$ was registered: then one has
\ban
\moy{A}_\psi&=&\lim_{N\rightarrow\infty}\frac{1}{N}\sum_{i=1}^N a^{(i)}\,=\,\sum_{k=1}^d \left(\lim_{N\rightarrow\infty}\frac{n_k}{N}\right)\,a_k\,=\, \sum_{k=1}^d P(\phi_k|\psi)\,a_k\,=\,\sandwich{\psi}{A}{\psi}\,.
\ean
It is however important to keep in mind that, strictly speaking, \textit{only the orthonormal basis defines an ideal measurement, while the labeling of the outcomes of a measurement is arbitrary}: even if real numbers are a very convenient choice in many cases, one can use complex numbers, vectors, colors or any other symbol. In other words, \textit{what is directly measured in an ideal measurement are the probabilities of each outcome}; average values are derived quantities. 

\subsubsection{Mixed states}

The simplest way of introducing mixed states is to think of a source that fluctuates, so that it produces the pure state $\ket{\psi_1}$ with probability $p_1$, $\ket{\psi_2}$ with probability $p_2$ and so on. If nothing is known about these fluctuations, the statistics of any ideal measurement will look like
\ban
P(\phi_j|\rho)\,=\,\sum_{k}p_k\,P(\phi_j|\psi_k)\,=\,\Tr\left(P_{\phi_j}\rho\right)&\mathrm{ with }& \rho=\sum_{k}p_kP_{\psi_k}\,.
\ean
This is the usual definition of the \textit{density matrix}. It is again important to stress that the notion of mixed state is not proper to quantum physics: in fact, in the formula above, only the $P(\phi_j|\psi_k)$ are typically quantum, but the probabilities $p_k$ are classical (at least formally; their ultimate origin may be entanglement, see below, but this is another matter).

\subsubsection{A remark on Gleason's theorem}

At this stage, it is worth while mentioning Gleason's theorem \cite{gleason}. Suppose that properties $P$ are defined by sub-spaces ${\mathcal E}_P$ of a Hilbert space ${\mathcal H}$; and suppose in addition that orthogonal sub-spaces are associated to distinguishable properties. Let $\omega$ be an assignment of probabilities, i.e. $\omega({\mathcal E}_P)\in[0,1]$, $\omega({\mathcal H})=1$ and $\omega({\mathcal E}_P\oplus {\mathcal E}_{P'})=\omega({\mathcal E}_P)+\omega({\mathcal E}_{P'})$ if ${\mathcal E}_P\perp {\mathcal E}_{P'}$. Then, if the dimension of the Hilbert space is $d\geq 3$, to each such $\omega$ one can associate a non-negative Hermitian operator $\rho_{\omega}$ such that $\omega({\mathcal E}_P)=\Tr(\rho_{\omega}\Pi_P)$ with $\Pi_P$ the projector on the subspace ${\mathcal E}_P$.

This complicated statement basically says that Born's rule can be \textit{derived} if one decides to associate properties to sub-spaces of a Hilbert space and to identify orthogonality with distinguishability. Curiously enough, if one allows for generalized measurements (see below), the theorem becomes valid for $d=2$ and the proof is considerably simplified \cite{fuchs}.

\subsection{Dynamics of closed systems: reversibility}

\subsubsection{Why reversible evolution must be unitary}

The requirement of reversible dynamics for closed systems implies that \textit{the evolution operator must be unitary}. As a simple way of understanding this, let us take an analogy with classical computation, that can be seen as an evolution of an initial string of bits. A computation is obviously reversible if and only if it consists in a permutation: indeed, a permutation is reversible; any other operation, that would map two strings onto the same one, is clearly not reversible. A reversible evolution in a vector space is slightly more general: it is \textit{a change of basis}. A change of basis is obviously reversible; the brute force proof of the converse is hard, but we can provide a proof ``by consistency of the theory'' that is quite instructive.

We are first going to argue that any operation that describes \textit{reversible} evolution in time must preserve the modulus of the scalar product, $\chi=|\braket{\psi_1}{\psi_2}|$, between any two vectors. The point is that, as a direct consequence of Born's rule for probabilities, $\chi$ has a crucial operational interpretation in the theory: two states can be perfectly discriminated if and only if $\chi=0$, and are identical if and only if $\chi=1$; this indicates (and it can be proved rigorously, see Section \ref{secstatediscr}) that $\chi$ quantifies the ``distinguishability'' between the two states in the best possible measurement. At this stage, we have only spoken of preparation and measurement and invoked Born's rule; so we have made no assumption on time evolution. So now suppose that an evolution in time can change $\chi$: if $\chi$ decreases, then the states become more distinguishable, contradicting the fact that $\chi$ quantifies the maximal distinguishability (in other words, in this case the best measurement would consist in waiting for some time before performing the measurement itself). So, for our measurement theory to be meaningful, $\chi$ can only increase. But if the evolution is reversible, by reverting it we would have a valid evolution in which $\chi$ decreases. The only remaining option is that $\chi$ does not change during reversible evolution. Unitarity follows from the fact that only unitary and anti-unitary operations preserve $\chi$, but anti-unitary operations cannot be used for a symmetry with a continuous parameter as translation in time is.

\subsubsection{The status of the Schr\"odinger equation}

If the generator of the evolution is supposed to be independent of time, standard group theoretical arguments and correspondence with classical dynamics allow writing $U=e^{-iHt/\hbar}$, where $H$ is the Hamilton operator; from this expression, the corresponding differential equation $i\hbar\frac{d}{dt}\ket{\psi}=H\ket{\psi}$ (Schr\"odinger equation) is readily derived. For the case where the dynamics itself varies with time, there is no such derivation: rather, one \textit{assumes} the same Schr\"odinger equation to hold with $H=H(t)$. Needless to say, the corresponding dynamics remains unitary.

\section{Composite systems}

In the classical case, as we have argued above, a pure state of a composite system is always a \textit{product state}, i.e. a state of the form $s_A\times s_B$, where $s_j$ is a pure state of system $j$; in the space of states, this translates by the Cartesian product of the sets of properties of each system.

In quantum theory, product states obviously exist as well, as they correspond to possible physical situations. But the space of states of the whole system cannot be simply ${\mathcal H}_A\times {\mathcal H}_B$, because this is not a vector space. The vector space that contains all product states and their linear combinations is the \textit{tensor product}
\ban
{\mathcal H}&=&{\mathcal H}_A\otimes {\mathcal H}_B\,.
\ean
We start with a rapid survey of the algebraic structure of the tensor product --- it really behaves like a product.

\subsection{Tensor product algebra}

The space ${\mathcal H}={\mathcal H}_A\otimes {\mathcal H}_B$ is constructively defined out of its components by requiring that: for any basis $\{\ket{\alpha_j}\}_{j=1...d_A}$ of ${\mathcal H}_A$ and for any basis $\{\ket{\beta_k}\}_{j=1...d_B}$ of ${\mathcal H}_B$, the set of states $\left\{\ket{\alpha_j}\otimes\ket{\beta_k}\right\}_{j,k=...}$ forms a basis of ${\mathcal H}$. It follows immediately that the dimension of ${\mathcal H}$ is $d_Ad_B$, as it should\footnote{Indeed, one can do the counting on product states: two product states of the composite system are perfectly distinguishable if either the two states of the first system are perfectly distinguishable, or the two states of the second system are perfectly distinguishable, or both.}.

The scalar product is defined on product states as
\ban
\left(\bra{\psi}\otimes\bra{\phi}\right)\left(\ket{\psi'}\otimes\ket{\phi'}\right) &=& \braket{\psi}{\psi'}\,\braket{\phi}{\phi'}
\ean and extended to the whole space by linearity.

The space of linear operators on ${\mathcal H}$ corresponds with the tensor product of the spaces of linear operators on ${\mathcal H}_A$ and on ${\mathcal H}_B$. An operator of the product form, $A\otimes B$, acts on product states as
\ban
\left(A\otimes B\right)\left(\ket{\psi}\otimes\ket{\phi}\right) &=& A\ket{\psi}\otimes B\ket{\phi}\,;
\ean
by linearity, this defines uniquely the action of the most general operator on the most general state.

\subsection{Entanglement}

Any linear combination of product states defines a possible pure state of the composite system. However, a state like
\ba
\ket{\psi(\theta)}&=&\cos\theta \ket{\alpha_1}\otimes\ket{\beta_1}+\sin\theta \ket{\alpha_2}\otimes\ket{\beta_2}
\ea
cannot be written as a product state, as it is easily checked by writing down the most general product state $\big(\sum_j a_j\ket{\alpha_j}\big)\otimes \big(\sum_k b_k\ket{\beta_k}\big)$ and equating the coefficients. In fact, it's easy to convince oneself that product states form a set of measure zero (according to any reasonable measure) in the whole set of states. \textit{A pure state, that cannot be written as a product state, is called ``entangled''}.

The most astonishing feature of entanglement is that we have a pure state of the composite system that does not arise from pure states of the components: in other words, the properties of the whole are sharply defined, while the properties of the sub-systems are not.

\subsubsection{Case study: the singlet state}

We are going to study a specific example that plays an important role in what follows. Like most of the examples in this series of lectures, this one involves \textit{qubits}, i.e. two-level systems. The algebra of two-level systems is supposed to be known from basic quantum physics; for convenience, it is reminded as an Appendix to this lecture (section \ref{qubits}).

Consider two qubits. The state
\ba
\ket{\Psi^-}&=&\frac{1}{\sqrt{2}}\left(\ket{0}\otimes\ket{1}-\ket{1}\otimes\ket{0}\right)\label{singlet}
\ea is called \textit{singlet} because of its particular status in the theory of addition of angular momenta. The projector on this state reads
\ba
P_{\Psi^-}&=&\frac{1}{4}\big(\one\otimes\one-\sigma_x\otimes\sigma_x -\sigma_y\otimes\sigma_y -\sigma_z\otimes\sigma_z\big)\,.
\ea
It appears clearly from the projector\footnote{For those who have never done it, it is useful to check that indeed the state has the same form in any basis. To do so, write a formal singlet in a different basis $\frac{1}{\sqrt{2}}\left(\ket{+\hat{n}} \otimes\ket{-\hat{n}}+\ket{-\hat{n}}\otimes\ket{+\hat{n}}\right)$, open the expression up using (\ref{spinstate}) and verify that this state is exactly the same state as (\ref{singlet}).} that this state is \textit{invariant by bilateral rotation}: $u\otimes u\ket{\Psi^-}=\ket{\Psi^-}$ (possibly up to a global phase). In particular, consider a measurement in which the first qubit is measured along direction $\vec{a}$ and the second along direction $\vec{b}$. The statistics of the outcomes $r_A,r_B\in\{-1,+1\}$ are given by $P_{\vec{a},\vec{b}}(r_A,r_B)=\left|\left(\bra{r_A\vec{a}}\otimes\bra{r_B\vec{b}}\right)\ket{\Psi^-}\right|^2$. The calculation leads to
\ba
P_{\vec{a},\vec{b}}(++)\,=\,P_{\vec{a},\vec{b}}(--)&=&\frac{1}{4}\big(1-\vec{a}\otimes\vec{b}\big)\,,\\
P_{\vec{a},\vec{b}}(+-)\,=\,P_{\vec{a},\vec{b}}(-+)&=&\frac{1}{4}\big(1+\vec{a}\otimes\vec{b}\big)\,.
\ea
Much information can be extracted from those statistics, we shall come back to them in Lecture 4. For the moment, let us just stress the following features:
\begin{itemize}
\item On the one hand, $P_{\vec{a},\vec{b}}(r_A=+)=P_{\vec{a},\vec{b}}(r_A=-)=\demi$ and $P_{\vec{a},\vec{b}}(r_B=+)=P_{\vec{a},\vec{b}}(r_B=-)=\demi$: the outcomes of the measurements of each qubit appear completely random, independent of the directions of the measurements. This implies the operational meaning of the fact that the state of each qubit cannot be pure (and here it is actually maximally mixed, see below).
\item On the other hand, there are very sharp correlations: in particular, whenever $\vec{a}=\vec{b}$, the two outcomes are rigorously opposite.
\end{itemize} 
This can be seen as defining the relation ``being opposite'' for two arrows in itself, without specifying in which direction each arrow points. While this possibility is logically compelling, it is worth while reminding that this is impossible in our everyday life.

A warning: we stressed that the singlet has a special role to play in the addition of two spins 1/2: indeed, it defines a one-dimensional subspace of total spin 0, as opposite to the three states that form the spin-1 triplet. However, in quantum information theory this feature is \textit{not} crucial. In other words, in quantum information the state $\ket{\Psi^-}$ is as good as any other state $u_1\otimes u_2\ket{\Psi^-}=\frac{1}{\sqrt{2}}\left(\ket{+\hat{m}}\otimes\ket{+\hat{n}}+\ket{-\hat{m}}\otimes\ket{-\hat{n}}\right)$ that can obtained from it with local unitaries. In particular, it is customary to define the states
\ba
\ket{\Psi^+}&=&\frac{1}{\sqrt{2}}\left(\ket{0}\otimes\ket{1}+\ket{1}\otimes\ket{0}\right)\,,\\
\ket{\Phi^+}&=&\frac{1}{\sqrt{2}}\left(\ket{0}\otimes\ket{0}+\ket{1}\otimes\ket{1}\right)\,,\\
\ket{\Phi^-}&=&\frac{1}{\sqrt{2}}\left(\ket{0}\otimes\ket{0}-\ket{1}\otimes\ket{1}\right)
\ea that are orthogonal to $\ket{\Psi^-}$ and with which they form the so-called \textit{Bell basis}.

\subsubsection{Entanglement for mixed states}

It is in principle not difficult to decide whether a pure state is
entangled or not: just check if it is a product state; if not,
then it is entangled. For mixed states, the definition is more
subtle, because a mixed state may exhibit classical correlations.
As an example, consider the mixture ``half of the times I prepare
$\ket{0}\otimes\ket{0}$, and half of the times I prepare
$\ket{1}\otimes\ket{1}$'', that is,
$\rho=\demi\ket{00}\bra{00}+\demi \ket{11}\bra{11}$ (with obvious
shortcut notations). This state exhibits correlations: whenever both
particles are measured in the basis defined by $\ket{0}$ and
$\ket{1}$, the outcomes of both measurements are the same.
However, by the very way the state was prepared, it is clear that
no entanglement is involved.

We shall then call {\em separable} a mixed state for which a
decomposition as a convex sum of product state exists: \ba
\rho&=&\sum_k\,p_k\,\big(\ket{\psi_k}\bra{\psi_k}\big)_A\otimes
\big(\ket{\phi_k}\bra{\phi_k}\big)_B\,. \ea It is enough that a single
mixture\footnote{Recall that any mixed state that is not pure can
be decomposed in an infinite number of ways as a mixture of pure
states (although generally there is only one mixture of pure {\em
mutually orthogonal} states, because of hermiticity).} of product
states exists, for the state to be separable. A mixed state is called {\em entangled} if it is not separable, i.e. if some entangled state must be used in order to prepare it \cite{wer}. Apart from some
special cases, to date no general criterion is known to decide
whether a given mixed state is separable or entangled.

\subsection{Partial states, no-signaling and purification}

We have just seen that, in the presence of entanglement, one
cannot assign a separate pure state to each of the sub-systems.
Still, suppose two entangled particles are sent apart from one another to two physicists, Alice and Bob. Alice on her location holds particle A and can make measurements on it, without possibly even knowing that a guy named
Bob holds a particle that is correlated with hers. Quantum physics
should provide Alice with rules to compute the probabilities for
the outcome she observes: there must
be a ``state'' (positive, unit trace hermitian operator) that
describes the information available on Alice's side. Such a state
is called ``partial state'' or ``local state''. The local state cannot
be pure if the state of the global system is entangled.

Suppose that the full state is $\rho_{AB}$. The partial state
$\rho_A$ is defined as follows: for any observable ${\mathcal A}$ on Alice's
particle, it must hold \ba \mbox{Tr}_A\big(\rho_A\,{\mathcal A}\big) &=&
\mbox{Tr}_{AB}\big(\rho_{AB}\,({\mathcal A}\otimes\one_B)\big)\,=\\&=&
\sum_{j=1}^{d_A}\sum_{k=1}^{d_B}\sandwich{a_j,b_k}{\rho_{AB}\,
({\mathcal A}\otimes\one_B)}{a_j,b_k}\,=\nonumber\\ &=&
\sum_{j=1}^{d_A}\sandwich{a_j}{\big(\mbox{Tr}_B(\rho_{AB})\big)\,{\mathcal A}}{a_j}
\,=\, \mbox{Tr}_A\big(\mbox{Tr}_B(\rho_{AB})\,{\mathcal A}\big)\nonumber \ea
so by identification \ba \rho_A&=& \mbox{Tr}_B(\rho_{AB})\,=\,
\sum_{k=1}^{d_B}\sandwich{b_k}{\rho_{AB}}{b_k}\,.
\label{partialgen}\ea This is the general definition of the
partial state: the partial state on A is obtained by \textit{partial
trace} over the other system B. The result (\ref{partialgen}) is
not ambiguous since the trace is a unitary invariant, that is,
gives the same result for any choice of basis. This implies that
{\em whatever Bob does, the partial state of Alice will remain
unchanged}. This fact has an important consequence, namely that Bob cannot use
entanglement to send any message to Alice. This is known as the
principle of {\em no-signaling through entanglement}\footnote{This
principle is sometimes stated by saying that entanglement does not
allow to signal ``faster-than-light''. Indeed, if entanglement would
allow signaling, this signal would travel faster than light; and
this is why we feel relieved when we notice that quantum physics
does not allow such a signaling. However, entanglement does not
allow to signal {\em tout court}, be it faster or slower than light.}.

When it comes to computing partial states, one can always come back to the general definition (\ref{partialgen}), but there are often more direct ways. For instance, no-signaling itself can be used: to obtain (say) $\rho_A$, one may consider that Bob performs a given measurement and compute the mixture that would be correspondingly prepared on Alice's side. This mixture must be $\rho_A$ itself, since the partial state is the same for any of Bob's actions.

\subsubsection{The notion of ``purification''}

The notion of \textit{purification} is, in a sense the reverse, of the notion of partial trace. It amounts at the following: any mixed state can be seen as the partial state of a pure state in a larger Hilbert space. To see that this is the case, note that any state $\rho$ can be diagonalized, i.e. there exist a set of orthogonal vectors $\{\ket{\varphi_k}\}_{k=1...r}$ ($r$ being the rank of $\rho$) such that $\rho=\sum_{k=1}^r p_k\,\ket{\varphi_k}\bra{\varphi_k}$. Then one can always take $r$ orthogonal vectors $\{\ket{e_k}\}_{k=1...r}$ of an ancilla and construct the pure state $\ket{\Psi}=\sum_{k=1}^r\sqrt{p_k}\ket{\varphi_k}\otimes\ket{e_k}$, of which $\rho$ is the partial state for the first sub-system.

This construction shows that a purification always exists; but because the decomposition of a mixed state onto pure states is not unique, the purification is not unique as well. In particular, if $\ket{\Psi}$ is a purification of $\rho$ and $U$ is a unitary on the ancilla, then $\one\otimes U\ket{\Psi}$ is also a purification of $\rho$, because it just amounts at choosing another set $\{\ket{e'_k}\}_{k=1...r}$ of orthogonal vectors. Remarkably though, this is the only freedom: it can indeed be proved that all purification are equivalent up to local operations on the ancilla. In this sense, the purification can be said to be unique.

\subsection{Measurement and evolution revisited}

To conclude this first Lecture, we have to mention one of the most important extensions of the notion of tensor product, namely the definitions of \textit{generalized measurements and evolution}. I shall not go into details of the formalism; whenever these notions appear later in these lectures, it will always be in a rather natural way.

\subsubsection{Evolution: CP maps, decoherence}

We have stressed above that the evolution of a closed system is reversible, hence unitary. In general, however, it is impossible to have a perfectly isolated system: even if some degrees of freedom (``the system'') are carefully selected and prepared in a given state, the evolution may imply some interaction with other degrees of freedom (``the environment''). In principle, it is obvious what has to be done to describe such a situation: consider both the system and the environment, study the unitary evolution, then trace the environment out and study the resulting state of the system. By doing this study for any given time $t$, one defines an effective map $T_t$ such that $\rho_S(t)=T_t[\rho_S(0)]$. It turns out that the maps defined this way coincide with the family of \textit{trace-preserving, completely positive (CP) maps}. Let us comment on each of these important properties:
\begin{itemize}
\item A map is positive if it transforms positive operators into positive operators; in our case, this means that the density matrix will never develop any negative eigenvalue;
\item The trace-preserving property is self-explained: $\Tr[\rho_S(t)]=1$ for all $t$. Together with the previous, it implies that a density matrix remains a density matrix --- an obvious necessity.
\item Not all positive maps, however, define possible evolutions of a sub-system. For instance, time-reversal is a positive map, but if one applies time-reversal only to a sub-system, it seems clear that one may get into trouble. A map $T$ is called ``completely positive'' if $T\otimes \one$ is also a positive map for any possible enlarged system. Interestingly, non-CP maps play an important role in quantum information: since the non-positivity can appear only if the system is entangled with the environment, these maps act as entanglement witnesses (see series of lectures on entanglement theory).
\end{itemize}

Trace-preserving CP maps are the most general evolution that a quantum system can undergo. Whenever the map on the system is not unitary, the state of the system becomes mixed by entanglement with the environment: this is called \textit{decoherence}. An obvious property of these maps is that states can only become ``less distinguishable''. A last remark: it would of course be desirable to derive a differential equation for the state of the system, whose solution implements the CP-map, without having to study the environment fully. In full generality, this task has been elusive; a general form has been derived by Lindblad for the case where the environment has no memory \cite{lindblad}. For a detailed introduction to the theory of open quantum systems, we refer to the book by Breuer and Petruccione \cite{petru}.

\subsubsection{Generalized measurements}

For measurements, a similar discussion can be made. The most general measurement on a quantum system consists in appending other degrees of freedom, then performing a measurement on the enlarged system. These additional degrees of freedom are called \textit{ancillae} (latin for ``servant maids'') rather than environment, but play exactly the same role: the final measurement may project on states, in which the system and the ancillae are entangled. Note that the ancillae start in a state that is independent of the state of the system: they are part of the measuring apparatus, and the whole measurement can truly be said to be a measurement on the system alone.

The effective result on the system is captured by a family of positive operators: for all possible generalized measurement with $D$ outcomes, there exist a family $\{A_k\}_{k=1,...,D}$ of positive operators, such that $\sum_kA_k^{\dagger}A_k=\one$. If the system has been prepared in the state $\rho$, the probability of obtaining outcome $k$ is given by $p_k=\Tr(A_k\rho A_k^\dagger)$ and the state is subsequently prepared as $\rho_k=\frac{1}{p_k}A_k\rho A_k^\dagger$. Note that $D$ must be at most the dimension of the total Hilbert space ``system+ancillae'', but can of course be much larger than the dimension of the Hilbert space of the system alone.

For some reason, the name that has been retained for such generalized measurements is \textit{positive-operator-valued measurements, or POVMs}. Some specific examples will be presented in the coming lectures.

\section{Appendix: one-qubit algebra}
\label{qubits}

Since most of the examples in these lectures will be done on spin 1/2 systems (\textit{qubits}), I remind here for convenience some basic elements of the Hilbert space ${\mathcal H}=\compl^2$ and of the linear operators on that space.

We start by recalling the Pauli matrices: \ba
\si_x\,=\,\left(\begin{array}{cc} 0 & 1\\ 1 &
0\end{array}\right)\,,&\, \si_y\,=\,\left(\begin{array}{cc} 0 &
-i\\ i & 0\end{array}\right)\,,&\,
\si_z\,=\,\left(\begin{array}{cc} 1 & 0\\ 0 &
-1\end{array}\right)\,.\ea The computational basis $\big\{\ket{0},\ket{1}\big\}$ is universally assumed to be the
eigenbasis of $\si_z$, so that: \ba \si_z\ket{0}\,=\,\ket{0}\,&,&\, \si_z\ket{1}\,=\,-\ket{1}\,,\\
\si_x\ket{0}\,=\,\ket{1}\,&,&\, \si_x\ket{1}\,=\,\ket{0}\,,\\
\si_y\ket{0}\,=\,i\ket{1}\,&,&\, \si_y\ket{1}\,=\,-i\ket{0}\,. \ea
It is useful sometimes to write the Pauli matrices as \ba
\si_x\,=\,\ket{0}\bra{1}+\ket{1}\bra{0}\,,&\,
\si_y\,=\,-i\ket{0}\bra{1}+i\ket{1}\bra{0}\,,&\,
\si_z\,=\,\ket{0}\bra{0}-\ket{1}\bra{1}\,.\ea One has
$\mbox{Tr}(\si_k)=0$ and $\si_k^2=\one$ for $k=x,y,z$; moreover,
\ba \si_x\si_y\,=\,-\si_y\si_x \,=\, i\si_z\,&&\;\mbox{+ cyclic
permutations}\label{commut}\,. \ea

The generic {\em pure state} of a qubit is written $\ket{\psi}=\alpha\ket{0}\,+\,\beta\ket{1}$. The associated
projector is therefore \ba \ket{\psi}\bra{\psi}&=& |\alpha|^2
\ket{0}\bra{0}\,+\,
|\beta|^2\ket{1}\bra{1}\,+\,\alpha\beta^*\ket{0}\bra{1}
 \,+\,\alpha^*\beta\ket{1}\bra{0}\,= \\&=&\demi\, \Big(\one\, +\, \big(|\alpha|^2-
|\beta|^2\big)\,\si_z\,+\,2\mbox{Re}(\alpha\beta^*)\,\si_x\, +\,
2\mbox{Im}(\alpha\beta^*)\,\si_y \Big)\,. \ea It follows
immediately from $\si_k^2=\one$ that
$\ket{\psi}\bra{\psi}=\demi\left(\one+\sum_k
\moy{\si_k}_{\psi}\,\si_k\right)$. One writes \ba
\ket{\psi}\bra{\psi}\,=\, \demi\, \Big(\one\, +\,
\hat{n}\cdot\vec{\si} \Big)\,,&\mbox{ with
}&\,\hat{n}\,=\,\left(\begin{array}{c} \moy{\si_x}_{\psi}\\
\moy{\si_y}_{\psi}\\ \moy{\si_z}_{\psi}\end{array}\right) \,=\,
\left(\begin{array}{c} 2\mbox{Re}(\alpha\beta^*)\\
2\mbox{Im}(\alpha\beta^*)\\
|\alpha|^2- |\beta|^2\end{array}\right)\,. \label{proj1}\ea The
vector $\hat{n}$ is called Bloch vector, it corresponds to the
expectation value of the "magnetic moment" $\vec{\si}$ in the
given state. For pure states (the case we are considering here),
its norm is one. It is actually well-known that such vectors cover
the unit sphere (called the {\em Bloch sphere}, or the Poincar\'e
sphere if the two-level system is the polarization of light). In
fact, there is a one-to-one correspondence between unit vectors
and pure states of a two-level system given by the following
parametrization in spherical coordinates: \ba
\ket{\psi}\,\equiv\,\ket{+\hat{n}}\,=\,\cos\frac{\theta}{2}\,\ket{0}\,+\,
e^{i\varphi}\sin\frac{\theta}{2}\,\ket{1}& \leftrightarrow &
\hat{n}\equiv \hat{n}(\theta,\varphi)\,=\,\left(\begin{array}{c} \sin\theta\,\cos\varphi\\
\sin\theta\,\sin\varphi\\
\cos\theta \end{array}\right)\,. \label{spinstate}\ea In turn, $\ket{+\hat{n}}$ is
the eigenstate of $\hat{n}\cdot\vec{\si}$ for the eigenvalue $+1$;
or alternatively, \ba\hat{n}\cdot\vec{\si}\, \equiv\,\si_n&=&
\ket{+\hat{n}}\bra{+\hat{n}}- \ket{-\hat{n}}\bra{-\hat{n}}\,.\ea
One has $\mbox{Tr}(\si_n)=0$ and $\si_n^2=\one$. For any basis
$\big\{\ket{+\hat{n}},\ket{-\hat{n}}\big\}$, one has the closure (completeness) relation $\ket{+\hat{n}}\bra{+\hat{n}}+
\ket{-\hat{n}}\bra{-\hat{n}}=\one$. The eigenstates of $\si_x$ and $\si_y$ are frequently used. They read, up to a global phase:\ba\ket{\pm x}\,=\, \frac{1}{\sqrt{2}}\big(\ket{0} \pm \ket{1}\big)&,&\ket{\pm y} \,=\,\frac{1}{\sqrt{2}}\big(\ket{0} \pm i\ket{1}\big)\,.\ea

Let's move to the study of {\em mixed states}. Given that any
projector can be written as (\ref{proj1}), obviously any mixed
state takes exactly the same form. In fact, consider just the
mixture of two pure states: \ban \rho &=&
p_1\ket{\psi_1}\bra{\psi_1}+p_2 \ket{\psi_2}\bra{\psi_2}\,=\,
\demi\, \Big(\one\, +\, \big(
p_1\hat{n_1}+p_2\hat{n_2}\big)\cdot\vec{\si} \Big)\,. \ean It is
easy to verify that the resulting Bloch vector
$\vec{m}=p_1\hat{n_1}+p_2\hat{n_2}$ lies {\em inside} the unit
sphere. In fact, the points in the volume of the Bloch sphere are
in a one-to-one correspondance with all possible states of a
single qubit, be they pure (in which case their Bloch vector lies
on the surface) or mixed.

We can summarize all that should be known on the states of a
single qubit as follows: generically, a state of a single qubits
reads \ba \rho\,=\, \demi\, \Big(\one\, +\, \vec{m}\cdot\vec{\si}
\Big)\,=\, \demi\, \Big(\one\, +\, |\vec{m}|\si_m \Big)\,,&\mbox{
with
}&\,\vec{m}\,=\,\left(\begin{array}{c} \mbox{Tr}\big({\si_x}\rho\big)\\
\mbox{Tr}\big({\si_y}\rho\big)\\
\mbox{Tr}\big({\si_z}\rho\big)\end{array}\right)\,. \ea The norm
of the Bloch vector is $|\vec{m}|\leq 1$, with equality if and
only if the state is pure. Since $\rho$ is hermitian, there is one
and only one decomposition as the sum of two orthogonal
projectors. Clearly, the eigenstates of this decomposition are the
eigenstates $\ket{+\hat{m}}$ and $\ket{-\hat{m}}$ of
$\hat{m}\cdot\vec{\si}$, where $\hat{m}=\vec{m}/|\vec{m}|$. The
orthogonal decomposition reads \ba \rho &=&
\left(\frac{1+|\vec{m}|}{2}\right)\,\ket{+\hat{m}}\bra{+\hat{m}}\,+\,
\left(\frac{1-|\vec{m}|}{2}\right)\,\ket{-\hat{m}}\bra{-\hat{m}}\,.
\ea Recall that any density matrix $\rho$ that is not a projector
admits an infinity of decompositions as sum of projectors. All these decompositions are equivalent, because
the density matrix carries all the information on the state, that
is, on the actual properties of the system.

Finally, a useful result: the probability of finding
$\ket{+\hat{n}}$ given the state
$\rho=\demi(\one+\vec{m}\cdot\vec{\si})$ is \ba
\mbox{Prob}(+\hat{n}\,|+\vec{m}\,)&=&
\frac{1}{4}\,\mbox{Tr}\left[(\one+\hat{n}\cdot\vec{\si})
(\one+\vec{m}\cdot\vec{\si})\right]\,=\,\frac{1+\hat{n}\cdot\vec{m}}{2}
\,.\label{probaspin} \ea

\section{Tutorials}

\subsection{Problems}

\noindent\textbf{Exercise 1.1}

Which of the following states are entangled?
\begin{enumerate}
\item $\ket{\Psi_1}=\cos\theta\ket{0}\ket{0}+\sin\theta\ket{1}\ket{1}$.
\item $\ket{\Psi_2}=\cos\theta\ket{0}\ket{0}+\sin\theta\ket{1}\ket{0}$.
\item $\ket{\Psi_3}=\demi(\ket{0}\ket{0}+\ket{0}\ket{1}-\ket{1}\ket{0}-\ket{1}\ket{1})$.
\item $\ket{\Psi_4}=\demi(\ket{0}\ket{0}+\ket{0}\ket{1}+\ket{1}\ket{0}-\ket{1}\ket{1})$.
\end{enumerate}
For those that are not entangled, give the decomposition as a product state. For simplicity of notation, we write $\ket{\psi_1}\ket{\psi_2}$ instead of $\ket{\psi_1}\otimes\ket{\psi_2}$.\\

\noindent\textbf{Exercise 1.2}

Compute the partial states $\rho_A$ and $\rho_B$ for the following states of two qubits:
\begin{enumerate}
\item The pure state $\ket{\Psi}=\sqrt{\frac{2}{3}}\ket{0}\ket{+}\,+\,
\sqrt{\frac{1}{3}}\ket{+}\ket{-}$; where $\ket{\pm}=\frac{1}{\sqrt{2}}(\ket{0}\pm\ket{1})$.
\item The mixed state
$W(\lambda)\,=\,\lambda
\ket{\Psi^-}\bra{\Psi^-}+(1-\lambda)\frac{\one}{4}$, called Werner state [R.F. Werner, Phys. Rev. A {\bf 40}, 4277 (1989)].
\end{enumerate}
Verify in both cases that $\rho_A$ and $\rho_B$ are mixed by computing the norm of their Bloch vectors.\\

\noindent\textbf{Exercise 1.3}

We consider the following decoherent channel \cite{sca02}. A qubit, initially prepared in a state $\rho$, undergoes sequential ``collisions'' with qubits coming from a reservoir. All the qubits of the reservoir are supposed to be in state $\xi=p\ket{0}\bra{0}+(1-p)\ket{1}\bra{1}$. Each collision implements the evolution
\ba
U&:&\left\{\begin{array}{lcl}\ket{0}\ket{0}&\longrightarrow& \ket{0}\ket{0}\\
\ket{0}\ket{1}&\longrightarrow& \cos\phi\ket{0}\ket{1}+i\sin\phi\ket{1}\ket{0}\\
\ket{1}\ket{0}&\longrightarrow& \cos\phi\ket{1}\ket{0}+i\sin\phi\ket{0}\ket{1}\\
\ket{1}\ket{1}&\longrightarrow& \ket{1}\ket{1}\\
\end{array}\right.
\ea with $\sin\phi\neq 0$. We assume that each qubit of the reservoir interacts only once with the system qubit. Therefore, the state of the system after collision with $n+1$ qubits of the bath is defined recursively as \ba\rho^{(n+1)}\,\equiv\,T_\xi^{n+1}[\rho]&=&\Tr_{B}\left(U \rho^{(n)}\otimes \xi U^\dagger\right)\,.\label{cpmap}\ea
\begin{enumerate}
\item Let $\rho^{(n)}=d^{(n)}\ket{0}\bra{0}+(1-d^{(n)})\ket{1}\bra{1}+k^{(n)}\ket{0}\bra{1}+{k^{(n)}}^*\ket{1}\bra{0}$. Prove that the CP-map (\ref{cpmap}) induces the recursive relations
\ba
d^{(n+1)}\,=\,c^2d^{(n)}+s^2p&,&k^{(n+1)}\,=\,ck^{(n)}
\ea with $c=\cos\phi$ and $s=\sin\phi$.
\item By iteration, provide $d^{(n+1)}$ and $k^{(n+1)}$ as a function of the parameters of the initial state $d^{(0)}$ and $k^{(0)}$. Conclude that $T_\xi^{n}[\rho]\longrightarrow \xi$ when $n\rightarrow\infty$, whatever the initial state $\rho$ (pure or mixed).
\item We have just studied an example of ``thermalization'': a system, put in contact with a large reservoir, ultimately assumes the same state as the particles in the reservoir. Naively, one would have described this process as $\rho\otimes\xi^{\otimes N}\rightarrow \xi^{\otimes N+1}$ for all $\rho$. Why is such a process not allowed by quantum physics?
\item The condition $\sin\phi\neq 0$ is necessary to have a non-trivial evolution during each collision; however, to have a meaningful model of thermalization one has to enforce $\cos\phi>>|\sin\phi|$. What is the meaning of this condition? \textit{Hint}: as a counter-example, consider the extreme case $\sin\phi=1$: what is then $U$? What does the process look like in this case?
\end{enumerate}

\subsection{Solutions}

\noindent\textbf{Exercise 1.1}

$\ket{\Psi_1}$ is entangled. $\ket{\Psi_2}=(\cos\theta\ket{0}+\sin\theta\ket{1})\ket{0}$ is not entangled. $\ket{\Psi_3}=\demi(\ket{0}-\ket{1})(\ket{0}+\ket{1})=\ket{-}\ket{+}$ is not entangled. $\ket{\Psi_4}$ is entangled: this can be verified by direct calculation, or also by noticing that $\ket{\Psi_4}=\frac{1}{\sqrt{2}}(\ket{0}\ket{+}+\ket{1}\ket{-})$; by just relabeling the basis of the second system, one sees that this state has the same form as $\ket{\Psi_1}$ with $\cos\theta=\sin\theta=\frac{1}{\sqrt{2}}$.\\

\noindent\textbf{Exercise 1.2}

\begin{enumerate}
\item For the pure state under study,  $\rho_A=\frac{2}{3}\ket{0}\bra{0}+\frac{1}{3}\ket{+}\bra{+}=\demi\left(\one+\frac{2}{3}\sigma_z+\frac{1}{3}\sigma_x\right)$. In order to compute $\rho_B$, here is a possibility (maybe not the fastest one): first, rewrite the state as $\ket{\Psi}=\ket{0}\Big(\sqrt{\frac{2}{3}}\ket{+}+\sqrt{\frac{1}{6}}\ket{-}\Big)\,+\,
\sqrt{\frac{1}{6}}\ket{1}\ket{-}$. Then
\ban
\rho_B&=&\Big(\sqrt{\frac{2}{3}}\ket{+}+\sqrt{\frac{1}{6}}\ket{-}\Big)\Big(\sqrt{\frac{2}{3}}\bra{+}+\sqrt{\frac{1}{6}}\bra{-}\Big)\,+\,\frac{1}{6}\ket{-}\bra{-}\\
&=&\frac{2}{3}\ket{+}\bra{+}+\frac{1}{3}\ket{-}\bra{-}+\frac{1}{3}\ket{+}\bra{-}+\frac{1}{3}\ket{-}\bra{+}\,=\,\frac{2}{3}\ket{0}\bra{0}+\frac{1}{3}\ket{+}\bra{+}\,=\,\rho_A
\ean since $\ket{0}=\frac{1}{\sqrt{2}}(\ket{+}+\ket{-})$. Note how, in this last calculation, the normalization is taken care of automatically.

The Bloch vector of $\rho_A=\rho_B$ has norm $|\vec{m}|=\frac{\sqrt{5}}{3}<1$, therefore the states are mixed.

\item For the Werner state: $\rho_A=\rho_B=\frac{\one}{2}$. These states are maximally mixed, indeed $|\vec{m}|=0$.
\end{enumerate}

\noindent\textbf{Exercise 1.3}

\begin{enumerate}
\item The first point is a matter of patience in writing down explicitly $U \rho^{(n)}\otimes \xi U^\dagger$, then noticing that $\Tr(\ket{0}\bra{0})=\Tr(\ket{1}\bra{1})=1$ and $\Tr(\ket{0}\bra{1})=\Tr(\ket{1}\bra{0})=0$.
\item For the off-diagonal term, the recursion is obviously
\ban
k^{(n+1)}&=&c^{n+1}k^{(0)}\,.
\ean
For the diagonal term, one has
\ban
d^{(n+1)}&=&c^2\left[c^2d^{(n-1)}+s^2p\right]+s^2p\,=\,c^4d^{(n-1)}+s^2(1+c^2)p\,=\,...\\
&=&c^{2(n+1)}d^{(0)}+s^2\sum_{k=0}^nc^{2k}\,p\,=\,c^{2(n+1)}d^{(0)}+[1-c^{2(n+1)}]p
\ean because $\sum_{k=0}^nc^{2k}=\frac{1-c^{2(n+1)}}{1-c^{2}}=\frac{1-c^{2(n+1)}}{s^{2}}$. Therefore $d^{(n+1)}\rightarrow p$ and $k^{(n+1)}\rightarrow 0$ for $n\rightarrow\infty$.
\item The evolution $\rho\otimes\xi^{\otimes N}\rightarrow \xi^{\otimes N+1}$ is not unitary, since two initially different states would end up being the same.
\item For $\sin\phi=1$, $U$ is the swap operation. In this case, the ``thermalization'' would consist in dumping the initial system in the reservoir and replacing it with one of the qubits of the reservoir. Such a process would introduce a very large fluctuation in the reservoir. By setting $\cos\phi\approx 1$, on the contrary, one has $\Tr(\rho_j A)\approx \Tr(\xi A)$ for any qubit $j$, for any single-particle physical quantity $A$. In other words, the system \textit{appears} to be completely thermalized and one has to measure some multi-particle physical quantities to see some differences. This view is perfectly consistent with the idea that irreversibility is only apparent.
\end{enumerate}

\chapter{Primitives of quantum information (I)}
\label{chap2}

\section{A tentative list of primitives}

The main tasks of quantum information, at the present stage of its development, are quantum computing and quantum cryptography. These tasks are complex: they rely on simpler notions, most of which are of interest in themselves. These notions that subtend the whole field are those that I call ``primitives''. Here is my tentative list of primitives, listed in chronological order of their appearance in the development of quantum physics:
\begin{enumerate}
\item Violation of Bell's inequalities
\item Quantum cloning
\item State discrimination
\item Quantum coding
\item Teleportation
\item Error correction
\item Entanglement distillation
\end{enumerate}
The common feature of all these primitives is that they have been studied in great detail. This does not mean that there are no open issues left; however, with a few remarkable exceptions, those are generally difficult points of rather technical nature. This is why you may not hear many talks dedicated to these topics in research conferences --- but the notions are there and will appear over and over again, as something anyone should know. This is why it is important to review those basic notions in a school like this one.

In my series of lectures, I shall deal with quantum cloning, teleportation and entanglement distillation (this Lecture), state discrimination, quantum coding (Lecture 3) and the violation of Bell's inequalities in greater detail (Lectures 4-6). Error correction is presented in this school by other lecturers.

\section{Quantum cloning}

The first primitive that should be considered is \textit{quantum cloning}. The famous no-go theorem was formulated in 1982-83 \cite{zurek,dieks,milonni,mandel}; much later, in 1996, came the idea of studying optimal cloning \cite{hb}. Since then, the subject has been the object of rather thorough investigations; two very comprehensive review articles are available \cite{sca05,cerf}.

\subsection{The no-go theorem}

It is well-known that one cannot measure the state $\ket{\psi}$ of
a single quantum system: the result of any single measurement of
an observable $A$ is one of its eigenstates, unrelated to the
input state $\ket{\psi}$. To reconstruct $\ket{\psi}$ (or more
generally $\rho$) one has to measure the average values of several
observables; this implies a statistic over a large number of identically prepared systems (see Lecture 3).

One can imagine to circumvent the problem in the following way:
take the system in the unknown state $\ket{\psi}$ and let it
interact with $N$ other systems previously prepared in a blank
reference state $\ket{R}$, in order to obtain $N+1$ copies of the
initial state: \ba
\ket{\psi}\otimes\ket{R}\otimes\ket{R}...\otimes\ket{R}
&\stackrel{?}{\longrightarrow}
\ket{\psi}\otimes\ket{\psi}\otimes\ket{\psi}...\otimes\ket{\psi}\,.
\ea Such a procedure would allow one to determine the quantum
state of a single system, without even measuring it because one
could measure the $N$ new copies and let the original untouched.
The no-cloning theorem of quantum information formalizes the
suspicion that such a procedure is impossible:

\textit{No-cloning theorem}: no quantum operation exists that can
duplicate perfectly an unknown quantum state.

The theorem can be proved by considering the $1\rightarrow 2$
cloning. Suppose first that perfect cloning is
possible without any ancilla: this means that there exist
a unitary operation such that \ba
\ket{\mbox{in}(\psi)}\,\equiv\,\ket{\psi}\otimes\ket{R}
&\stackrel{?}{\longrightarrow}
\ket{\psi}\otimes\ket{\psi}\,\equiv\, \ket{\mbox{out}(\psi)}\,.
\ea But such an operation cannot be unitary, because it does not
preserve the scalar product: \ban
\braket{\mbox{in}(\psi)}{\mbox{in}(\varphi)}
\,=\,\braket{\psi}{\varphi} &\neq &
\braket{\mbox{out}(\psi)}{\mbox{out}(\varphi)}
\,=\,\braket{\psi}{\varphi}^2\,. \ean Now we have to prove that perfect cloning is impossible also for CP maps, the most general evolution. So let's add an ancilla (the "machine") and suppose that \ba
\ket{\psi}\otimes\ket{R}\otimes\ket{M}
&\stackrel{?}{\longrightarrow}
\ket{\psi}\otimes\ket{\psi}\otimes\ket{M(\psi)} \label{perf2}\ea
is unitary. The same type of proof as before can be done as in
Sect. 9-4 of Peres' book \cite{peres}; here we give a different
one, closer to the Wootters-Zurek proof \cite{zurek}. We suppose
that (\ref{perf2}) holds for two orthogonal states, labelled
$\ket{0}$ and $\ket{1}$: \ban \ket{0}\otimes\ket{R}\otimes\ket{M}
&\longrightarrow \ket{0}\otimes\ket{0}\otimes\ket{M(0)}\\
\ket{1}\otimes\ket{R}\otimes\ket{M} &\longrightarrow
\ket{1}\otimes\ket{1}\otimes\ket{M(1)}\,. \ean Because of
linearity (we omit tensor products) then: \ban
\big(\ket{0}+\ket{1}\big)\ket{R}\,\ket{M} &\longrightarrow&
\ket{00}\,\ket{M(0)}+ \ket{11}\,\ket{M(1)}\ean that cannot be
equal to $\big(\ket{0}+\ket{1}\big)
\big(\ket{0}+\ket{1}\big)\,\ket{M(0+1)}=
\big(\ket{00}+\ket{10}+\ket{01}+\ket{11}\big)\,\ket{M(0+1)}$. So
(\ref{perf2}) may hold for states of a basis, but cannot hold for
all states. Since a unitary evolution with an ancilla is the most
general evolution allowed for quantum systems, the proof of the
theorem is concluded.

\subsection{How no-cloning was actually discovered}

Sometimes one learns more from mistakes than from perfect thought. This is, in my opinion, the case with the paper that triggered the discovery of no-cloning \cite{herbert}. The author, Herbert, reasoned as follows. Consider two particles in the singlet state: one particle goes to Alice, the other to Bob. If Alice measures $\sigma_z$, she prepares effectively on Bob's side either $\ket{+z}$ or $\ket{-z}$, with equal probability: therefore Bob's local state is $\rho_z=\demi\ket{+z}\bra{+z}+\demi\ket{-z}\bra{-z}=\frac{1}{2}\one$. If Alice measures $\sigma_x$, she prepares effectively on Bob's side either $\ket{+x}$ or $\ket{-x}$, with equal probability: Bob's local state is now $\rho_x=\demi\ket{+x}\bra{+x}+\demi\ket{-x}\bra{-x}=\frac{1}{2}\one$, equal to $\rho_z$, as it should because of no-signaling.

But suppose now that Bob can make a perfect copy of his qubit. Now, if Alice measures $\sigma_z$, Bob ends up with either $\ket{+z}\ket{+z}$ or $\ket{-z}\ket{-z}$, with equal probability: therefore Bob's local state is $\rho_z=\demi\ket{+z+z}\bra{+z+z}+\demi\ket{-z-z}\bra{-z-z}$. A similar reasoning leads to the conclusion that, if Alice measures $\sigma_x$, Bob's local state is $\rho_x=\demi\ket{+x+x}\bra{+x+x}+\demi\ket{-x-x}\bra{-x-x}$. But now, $\rho_x\neq \rho_z$! Herbert suggested that he had discovered a method to send signals faster than light --- had he been a bit more careful, he would have discovered the no-cloning theorem.

Astonishingly enough, Herbert's paper was published. Both referees are known: the late Asher Peres explained he knew the paper was wrong, but guessed it was going to trigger interesting developments \cite{peresherbert}; GianCarlo Ghirardi pointed out the mistake in his report, which may therefore be the first ``proof'' of the no-cloning theorem. Later, Gisin re-considered Herbert's scheme and studied how the two-copy state must be modified in order for $\rho_z$ and $\rho_x$ to be equal after imperfect duplication; he obtained an upper bound for the fidelity of the copies, that can actually be reached \cite{gisin}. 

A final point is worth noting. The whole reasoning of Herbert implicitly assumes that something does change instantaneously on a particle upon measuring an entangled particle at a distance. This view is not shared by most physicists. The whole debate might have gone astray on discussions about ``collapse'', just as the vague reply of Bohr to the Einstein-Podolski-Rosen paper prevented people from defining local variables in a precise way. It is fortunate that people seemed to have learned the lesson, and the problem was immediately cast in an \textit{operational} way. This is the ``spirit'' that later lead to the rise of quantum information science.

\subsection{The notion of Quantum Cloning Machines}

Perfect cloning of an unknown quantum state is impossible; conversely, cloning of orthogonal states belonging to a known basis is trivially possible: simply measure in the basis, and produce as many copies as you like of the state you obtained. What is also possible, it to {\em swap} the state from one system to the other: $\ket{\psi}\ket{R}\rightarrow \ket{R}\ket{\psi}$ is unitary; one has then created a perfect image of the input state on the second system, at the price of destroying the initial one.

The notion of {\em Quantum Cloning Machines} (QCM) is a wide notion encompassing all possible intermediate cases. One needs a figure of merit. Here we focus on the \textit{single-copy fidelity}, called ``fidelity'' for short. For each copied system $j$, this is defined as $F_{j}=\sandwich{\psi}{\rho_j}{\psi}$ for the initial state $\ket{\psi}$.

Here are some intuitive statements:
\begin{itemize}
\item If {\em perfect} cloning of an unknown quantum state is
impossible, {\em imperfect} cloning should be possible. In
particular, there should be an operation that allows to copy
equally well, with a fidelity $F<1$, any unknown state. Any such
operation will imply a ``degradation'' of the state of the original
system.

\item The fidelity of the ``original'' and of the ``copy'' after the
cloning need not be the same; the better the copy, the more the
original is perturbed.

\item One can also consider cloning $N\rightarrow M=N+k$; if
$N\rightarrow \infty$, the fidelity of the final $M$ copies can be
arbitrarily close to $1$.

\item Also, one may be willing to copy only a subset of all states.
\end{itemize}

Along with the variety of possible approaches, some terminology has been created:

{\it Universal QCM}: copies equally well all the states;
non-universal QCM, called state-dependent, have been studied only
in some cases, mostly related to the attacks of the spy on some
cryptography protocols.

{\it Symmetric QCM}: the original(s) and the copie(s) have the
same fidelity.

{\it Optimal QCM}: for a given fidelity of the original(s) after
interaction, the fidelities of the copie(s) is the maximal one.

\subsection{Case study: universal symmetric QCM $1\rightarrow 2$ for qubits}

We study in detail the first example of a quantum cloning
machine, the universal symmetric QCM $1\rightarrow 2$ for qubits
found by Bu\v{z}ek and Hillery \cite{hb}. It is however
instructive to start by analyzing first some trivial cloning strategies.

\subsubsection{Trivial cloning}

Consider the following strategy: \textit{Let the prepared qubit fly
unperturbed, and produce a new qubit in a randomly chosen state,
say $\ket{0}$. Don't keep track of which qubit is which.}

Let us compute the single-copy fidelity. We detect one particle: the original one with probability $\demi$,
the new one with the same probability. Thus the average
single-copy fidelity is \ba F_{triv}&=&
\demi\times 1\,+\,\demi\times\Big(\frac{1}{4\pi}\int_0^{2\pi}d\varphi
\int_{-1}^{1}d(\cos\theta)
\sandwich{\psi}{P_{0}}{\psi}\Big)=\nonumber\\
&=&\demi\,+\,\demi\Big(\frac{1}{2}\int_{-1}^{1}
d(\cos\theta)\frac{1+\cos\theta}{2}\Big)\,=\,\frac{3}{4}\,. \ea
It is interesting to see that a fidelity of 75\% can be reached by
such an uninteresting strategy. In particular, this implies that one must show $F>\frac{3}{4}$ in order to demonstrate
non-trivial cloning.

Note that one can consider another trivial cloning strategy, namely: \textit{measure the state in an arbitrary basis and produce two copies of the outcome.} A similar calculation to the one above shows that this strategy leads to $F=\frac{2}{3}$ for the average single-copy fidelity and is therefore worse than the previous one.

\subsubsection{The Bu\v{z}ek-Hillery (B-H) QCM for qubits}

The Bu\v{z}ek-Hillery (B-H) QCM is a universal symmetric QCM for
$1\rightarrow 2$ qubits, that was soon afterwards proved to be the
optimal one. We give no derivation, but rather start from the
definition and verify all the properties {\em a posteriori}.

The B-H cloner uses {\em three qubits}: the original (A), the copy
(B) and an ancilla (C). For convention, B and C are initially set
in the state $\ket{0}$. Here is the action in the computational
basis of A: \ba \begin{array}{ccc} \ket{0}\ket{0}\ket{0}
&\rightarrow & \sqrt{\frac{2}{3}}\, \ket{0}\ket{0}\ket{0}\,+\,
\sqrt{\frac{1}{6}}\,
\big[\ket{0}\ket{1}+\ket{1}\ket{0}\big]\ket{1}\\
\ket{1}\ket{0}\ket{0} &\rightarrow & \sqrt{\frac{2}{3}}\,
\ket{1}\ket{1}\ket{1}\,+\, \sqrt{\frac{1}{6}}\,
\big[\ket{1}\ket{0}+\ket{0}\ket{1}\big]\ket{0}\end{array}\,.
\label{hbbase}\ea These two relations induce the following action
on the most general input state $\ket{\psi}=\alpha\ket{0}+
\beta\ket{1}$: \ba \ket{\psi}\ket{0}\ket{0} &\rightarrow &
\sqrt{\frac{2}{3}}\, \ket{\psi}\ket{\psi}\ket{\psi^*}\,+\,
\sqrt{\frac{1}{6}}\, \big[\ket{\psi}\ket{\psi^\perp}+
\ket{\psi^\perp}\ket{\psi}\big]\ket{{\psi^*}^\perp}\,.
\label{hbgen}\ea We have written
$\ket{\psi^\perp}=\beta^*\ket{0}-\alpha^*\ket{1}$,
$\ket{\psi^*}=\alpha^*\ket{0}+ \beta^*\ket{1}$; combining the two
definitions one finds that $\ket{{\psi^*}^\perp}=
\ket{{\psi^\perp}^*}$. Eq. (\ref{hbgen}) is the starting point for
the subsequent analysis.

The verification that (\ref{hbgen}) follows from
(\ref{hbbase}) is made as follows: from (\ref{hbbase}), because of
linearity, \ban \ket{\psi}\ket{0}\ket{0} &\rightarrow &
\alpha\sqrt{\frac{2}{3}}\ket{000}+\alpha
\sqrt{\frac{1}{6}}[\ket{011}+\ket{101}]+
\beta\sqrt{\frac{2}{3}}\ket{111}+\beta
\sqrt{\frac{1}{6}}[\ket{100}+\ket{010}]\,. \ean Then one writes
explicitly the r.h.s. of (\ref{hbgen}) and finds the same state.

\subsubsection{B-H: State of A and B}

From Eq. (\ref{hbgen}), one sees immediately that A and B can be
exchanged, and in addition, that the transformation has the same
coefficients for all input state $\ket{\psi}$. Thus, the B-H QCM
is symmetric and universal. Explicitly, the partial states are \ba
\rho_A\,=\,\rho_B &=&
\frac{2}{3}\ket{\psi}\bra{\psi}+\frac{1}{3}\frac{\one}{2} \,=\,
\frac{5}{6}\ket{\psi}\bra{\psi}+\frac{1}{6}
\ket{\psi^\perp}\bra{\psi^\perp} \,=\, \frac{1}{2}\,\left(\one +
\frac{2}{3}\hat{m}\cdot\vec{\si}\right)\,. \label{rhoahb}\ea From
the standpoint of A then, the B-H cloner "shrinks" the Bloch
vector by a factor $\frac{2}{3}$ without changing its direction.

For both the original and the copy, the B-H cloner gives the
fidelity \ba
F_A\,=F_B\,=\,\sandwich{\psi}{\rho_A}{\psi}&=&\frac{5}{6}\,. \ea
This is the optimal fidelity for a symmetric universal
$1\longrightarrow 2$ cloner of qubits, a statement that is not
evident and was proved in later papers \cite{gisin,bruss,gm}.

\subsubsection{B-H: State of C}

Although it is a departure from the main theme, I find it
interesting to spend some words about the state of the ancilla C
after cloning. No condition has been imposed on this, but it turns
out to have a quite interesting meaning. We have \ba \rho_C &=&
\frac{2}{3}\ket{\psi^*}\bra{\psi^*}+\frac{1}{3}
\ket{{\psi^*}^\perp}\bra{{\psi^*}^\perp}\,=\,
\frac{1}{2}\,\left(\one +
\frac{1}{3}\hat{m}_*\cdot\vec{\si}\right) \label{rhochb}\ea with
$\hat{m}_*=(m_x,-m_y,m_z)$. This state is related to another
operation which, like cloning, is impossible to achieve perfectly,
namely the NOT operation that transforms
$\ket{\psi}=\alpha\ket{0}+ \beta\ket{1}$ into
$\ket{\psi^{\perp}}=\beta^*\ket{0}-\alpha^*\ket{1}$. Because of
the need for complex conjugation of the coefficients, the NOT
transformation is anti-unitary and cannot be
performed\footnote{Here is an intuitive version of this
impossibility result: any unitary operation on a qubit acts as a
rotation around an axis in the Bloch sphere, while the NOT is
achieved as the point symmetry of the Bloch sphere through its
center. Obviously, no rotation around an axis can implement a
point symmetry. A rotation of $\pi$ around the axis $z$ achieves
the NOT only for the states in the $(x,y)$ plane, while leaving
the eigenstates of $\si_z$ invariant.}. Just as for the cloning
theorem, one can choose to achieve the NOT on some states while
leaving other states unchanged; or one can find the operation that
approximates at best the NOT on all states, called the universal
NOT \cite{not}. This operation needs some ancilla, and reads
$\ket{\psi}\rightarrow \rho_{NOT} =
\frac{2}{3}\ket{\psi^\perp}\bra{\psi^\perp}+\frac{1}{3}
\ket{\psi}\bra{\psi}$. Now, it is easy to verify that
$\rho_{NOT}=\si_y\rho_C\si_y$ (just see the definition of
$\hat{m}_*$). Thus, the ancilla qubit of the B-H QCM carries (up
to a rotation of $\pi$ around the $y$ axis of the Bloch sphere)
the universal NOT of the input state.

\subsubsection{B-H as the coherent version of trivial cloning}

There is an intriguing link between the B-H QCM and the trivial cloning presented above. Recall that in the trivial cloning one has to ``forget'' which qubit is which in order to pick up one of the two qubits at random. In other words, the process involves summing over the classical permutations. One might ask what happens if this classical permutation is replaced by the \textit{quantum (coherent) permutation}. This operation is defined as the following CP-map:
\ba
T[\rho]&=&\frac{2}{3}\,S_2\, \big(\rho\otimes
\one \big)\,S_2 \label{clonwerner}
\ea
where $S_2$ is the projector on the symmetric space of two qubits, i.e. the 3-dimensional subspace spanned by $\{\ket{00},\ket{11},\ket{\Psi^+}\}$; the factor $\frac{2}{3}$ guarantees that the map is trace-preserving. Remarkably, for $\rho=\ket{\psi}\bra{\psi}$, $T[\rho]=\rho_{AB}(\psi)$ as obtained by applying the B-H QCM to $\ket{\psi}$. The map $T$ is not unitary, which justifies the need for the ancilla. This elegant construction was noted by Werner, who used it to find universal symmetric $N\rightarrow M$ cloning \cite{werner}.

\section{Teleportation}

The second primitive we are considering is \textit{teleportation}. Discovered over a black-board discussion, the teleportation protocol \cite{teleport} is a fascinating physical phenomenon (which is more, with a catchy name). Of course, contrary to quantum cloning, teleportation in itself admits few and rather obvious generalizations; in other words, ``teleportation'' is not and has never been a sub-field of quantum information. However, it is a seed for many other ideas and plays an important role in entanglement theory.

\subsection{The protocol}

As usual in these lectures, we present the protocol with qubits. Consider three qubits: qubit A is prepared in an arbitrary state $\ket{\psi}$; qubits B and C are prepared in a maximally entangled state, say $\ket{\Phi^+}$.

By writing $\ket{\psi}=\alpha\ket{0}+\beta\ket{1}$ and expanding the terms, one can readily verify
\ba
\ket{\psi}_A\ket{\Phi^+}_{BC}&=&\demi \Big[\ket{\Phi^+}_{AB}\ket{\psi}_C\,+\, \ket{\Phi^-}_{AB}\left(\sigma_z\ket{\psi}\right)_C\nonumber\\
&&+\, \ket{\Psi^+}_{AB}\left(\sigma_x\ket{\psi}\right)_C\,+\, \ket{\Psi^-}_{AB}\left(\tilde{\sigma}_y\ket{\psi}\right)_C\Big]
\label{idteleport}\ea
where $\tilde{\sigma}_y=-i{\sigma}_y$. This identity is the basis of the \textit{teleportation protocol}:
\begin{enumerate}
\item Prepare the three qubits as described above; bring qubits A and B together.
\item Perform a \textit{Bell-state measurement} on qubits A and B, and send the result of the measurement (2 bits) to the location of qubit C.
\item Upon reception of this information, apply the suitable unitary operation to C in order to recover $\ket{\psi}$.
\end{enumerate}
That is all for the protocol. Still, some remarks are worth making:
\begin{itemize}
\item It is customary to emphasize that \textit{information, not matter, is teleported:} qubit C had to exist in order to receive the state of qubit A. That being clarified, the name ``teleportation'' is well chosen: the information has been transferred from A to C without ever being available in the region when particle B has propagated.
\item Another point that is usually stressed is the fact that, of course, this task respects \textit{no-signaling}: indeed, the teleportation can be achieved only when the two bits of classical communication are sent, and these cannot travel faster than light. The phenomenon is nonetheless remarkable, because the two classical bits are definitely not sufficient to carry information about a state of a qubit (a vector in the Poincare sphere is defined by three continuous parameters).
\item Even though some of the particles may be ``propagating'', the qubit degree of freedom that is going to be teleported does not evolve in the protocol (there is no hamiltonian anywhere). Teleportation is due to the \textit{purely kinematical} identity (\ref{idteleport}).
\item Since the protocol can teleport every pure state with perfect fidelity, it can \textit{teleport any mixed state} as well.
\end{itemize}

\subsection{Entanglement swapping}

If the qubit to be teleported is itself part of an entangled pair, the identity (\ref{idteleport}) applied to B-CD leads to
\ba
\ket{\Phi^+}_{AB}\ket{\Phi^+}_{CD}&=&\demi \,\Big[\ket{\Phi^+}_{AD}\ket{\Phi^+}_{BC}\,+\, \ket{\Phi^-}_{AD}\ket{\Phi^-}_{BC}\nonumber\\
&&+\, \ket{\Psi^+}_{AD}\ket{\Psi^+}_{BC}+\ket{\Psi^-}_{AD}\ket{\Psi^-}_{BC}\Big]
\label{idswapping}\ea
because $\one\otimes\sigma_z\ket{\Phi^+}=\ket{\Phi^-}$, $\one\otimes\sigma_x\ket{\Phi^+}=\ket{\Psi^+}$ and $\one\otimes\tilde{\sigma}_y\ket{\Phi^+}=\ket{\Psi^-}$. Therefore, by performing the Bell-state measurement on B and C and sending the result to $D$, one can prepare the particles A and D in the state $\ket{\Phi^+}$, even if they have never interacted. This protocol is called \textit{entanglement swapping} \cite{yurke,zuk}.

Entanglement swapping, in itself, is nothing but a special case of teleportation, in which the particle to be teleported is itself part of an entangled pair. However, it shows that \textit{direct interaction is not needed to create entanglement}.

\subsection{Teleportation and entanglement swapping as primitives for other tasks}

\subsubsection{Two-qubit maximally entangled states as universal resources}

The possibility of teleportation has a fundamental consequence in entanglement theory, namely that bipartite maximally entangled states are universal resources to distribute entanglement. Indeed, suppose that $N$ partners want to share a fully $N$-partite entangled state. It is enough for \textit{one} of the partners, Paul, to act as a provider. Indeed, if Paul shares a bipartite maximally entangled state with each of the others, he can prepare locally the $N$-partite state, then teleport the state of each particle to the suitable person.

Now, maximally entangled states of any dimension can be created from enough many copies of two-qubit maximally entangled states (at least on paper), as the following example makes clear:
\ba
\ket{\Phi^+}_{AB}\ket{\Phi^+}_{A'B'}&=&\demi\big[\ket{00}_{AA'}\ket{00}_{BB'} + \ket{01}_{AA'}\ket{01}_{BB'} \nonumber\\&&+ \ket{10}_{AA'}\ket{10}_{BB'} + \ket{11}_{AA'}\ket{11}_{BB'}\big]\nonumber\\&=& \demi\big[\ket{0}_{\mathbf{A}}\ket{0}_{\mathbf{B}} + \ket{1}_{\mathbf{A}}\ket{1}_{\mathbf{B}} + \ket{2}_{\mathbf{A}}\ket{2}_{\mathbf{B}} + \ket{3}_{\mathbf{A}}\ket{3}_{\mathbf{B}}\big]\,.
\ea
In conclusion, as soon as the partners can share pairwise maximally entangled states of two-qubits with one provider, they can share any entangled state of arbitrary many parties and dimensions. Let me stress again that this is a theoretical statement about the behavior of entanglement as a resource, not necessarily a practical or even feasible scheme to realize an experiment.

\subsubsection{Quantum repeaters}

At the other end of the spectrum, the idea of quantum repeaters is triggered by a very practical problem. In quantum communication, one normally uses photons because they propagate well and interact little. In this case, decoherence is by far not the most problematic issue: \textit{losses} are. In other words, quantum communication schemes reach their limits when the photons just don't arrive often enough for the signal to overcome the noise of the detectors and other local apparatuses. Since you cannot amplify your signal because of the no-cloning theorem, losses seem unbeatable. Quantum repeaters are a clever solution to the problem \cite{repeaters,dlcz}. For exhaustive information, a recent review article on the topic is available \cite{revrep}.

Suppose you want to share an entangled pair between locations A and B, at a distance $\ell$. The transmission typically scales exponentially $t=10^{-\alpha \ell}$. What happens if A and B can both send one photon to an intermediate location C, where someone performs the Bell-state measurement? This does not seem to help, because the transmission from A to C is $\sqrt{t}$, and so is the transmission from B to C, so the probability that both photons arrive in C is still $t$. However, if C has \textit{quantum memories}, the picture changes significantly: now, C can establish a link with A and \textit{independently} a link with B.

This effect is best understood in terms of the time needed to establish an entangled pair between A and B. If photons have to travel from A to B, this time is (in suitable units) $\tau_{AB}^{(1)}=\frac{1}{t}$. In the same units, $\tau_{AC}^{(1)}=\tau_{BC}^{(1)}=\frac{1}{\sqrt{t}}$. If the two links can be established independently, it take to C in average $\demi\frac{1}{\sqrt{t}}$ to establish the first of the links, because the first can be either; conditioned now on the fact that one link was established, C has to wait in average $\frac{1}{\sqrt{t}}$ to establish the second link. After this, entanglement swapping can be performed, thus establishing the entangled pair between A and B in a time $\tau_{AB}^{(1)}\approx\frac{3}{2}\frac{1}{\sqrt{t}}$. For more details on this calculation, see Appendix B in \cite{revqkd}.

\section{Entanglement distillation}

This section is scandalously short. The reason is that I thought other lecturers would introduce the notion of distillation, but it turned out they had planned their lectures differently and there was no time for it! When we realized our lack of coordination, we decided that I would at least mention rapidly the idea, for the sake of completeness. So here it is. All the meaningful notions, extensions and references can be found for instance in the comprehensive review paper devoted to entanglement theory written by the Horodecki family \cite{H4}.

\subsection{The notion of distillation}

We have seen above that two-qubit maximally entangled states are a universal resource for distributing quantum states. Suppose now two parties Alice and Bob (it can of course be generalized) share many copies of some other, less entangled state $\rho_{AB}$: can they somehow ``concentrate'' or ``distill'' the entanglement they have, in order to end up with fewer copies of those very useful maximally entangled state? Of course, for the task to make sense, the distillation must be performed only using operations that themselves do not increase entanglement on a single-copy level: these are \textit{local operations and classical communication (LOCC)}. In other words, Alice and Bob are in different locations and they cannot send quantum systems to each other (otherwise trivially Alice would prepare a maximally entangled state and give or send one particle to Bob).

So, is it possible to achieve
\ba
{\rho_{AB}}^{\otimes N}&\stackrel{LOCC}{\longrightarrow}& \Phi^{\otimes m}\otimes\mathrm{(garbage)}
\ea
with $\Phi=\ket{\Phi}\bra{\Phi}$ a maximally entangled state of two qubits? The answer is: often yes but sometimes no! For instance, entanglement is distillable for all entangled pure states; also for all entangled states of two qubits, or of a qubit and a qutrit. Some entangled mixed state of higher dimensions and/or of more parties, however, are such that their entanglement is not distillable! In other words, you need entanglement to create them, but this entanglement cannot be recovered. Such states are called \textit{bound-entangled}.

At the moment of writing, the question of assessing whether or not an entangled state is distillable is still open in general. Those who want to have an idea of how many notions I am skipping here --- partial transpose (a positive, but non-completely-positive map), different measures of entanglement, etc. --- may browse the review paper mentioned above. But, in the context of a school, I cannot resist spelling out one of the most fascinating examples of bound-entangled states.

\subsection{A nice example of bound entangled state}

The example of bound entangled state that we are going to consider \cite{bennett} is a three-qubit state. Suppose that, for any reason, one starts out writing a basis of the Hilbert space as
\ban
\ket{\varphi_1}=\ket{0}\ket{1}\ket{+}\,,\; \ket{\varphi_2}=\ket{1}\ket{+}\ket{0}\,,\; \ket{\varphi_3}=\ket{+}\ket{0}\ket{1}\,,\; \ket{\varphi_4}=\ket{-}\ket{-}\ket{-}\,:\;
\ean
these states are obviously orthogonal, but four more states are needed to have a basis. Now, it turns out that the four remaining states cannot be product states: they must contain some entanglement! The four product states we started with form a so-called \textit{unextendible product basis}.

The next ingredient is the fact that the remaining four states can all be taken to be entangled only between the first two qubits; i.e. one can find four orthogonal vectors $\ket{\varphi_k(AB|C)}=\ket{\Psi_k}_{AB}\ket{\psi_k}_C$ for $k=5,6,7,8$ (I leave the interested reader to find out the explicit expressions). But of course, since the initial four states are symmetric under permutations, one might just as well choose the remaining $\ket{\varphi_k}$ to be of the form $\ket{\varphi_k(CA|B)}=\ket{\Psi_k}_{CA}\ket{\psi_k}_B$ or $\ket{\varphi_k(BC|A)}=\ket{\Psi_k}_{BC}\ket{\psi_k}_A$.

Consider now the state
\ba
\rho&=&\frac{1}{4} \left(\one-\sum_{k=1}^4\ket{\varphi_k}\bra{\varphi_k}\right)\,.
\ea
This is the maximally mixed state defined on the subspace that is complementary to the unextendible product basis. As such, it is obviously entangled, because there cannot be any product state in its support. But \textit{where} is the entanglement? Notice that
\ban
\rho&=&\frac{1}{4} \sum_{k=5}^8 \ket{\varphi_k(AB|C)}\bra{\varphi_k(AB|C)}\\&=& \frac{1}{4} \sum_{k=5}^8\ket{\varphi_k(CA|B)}\bra{\varphi_k(CA|B)}\\&=& \frac{1}{4} \sum_{k=5}^8\ket{\varphi_k(BC|A)}\bra{\varphi_k(BC|A)}\,:
\ean
therefore, according to \textit{each} possible bipartition, the state is separable. It is therefore impossible that separate parties can distill entanglement: $\rho$ is bound-entangled.

Is bound entanglement ``useful''? For long time, the answer was supposed to be negative. In 2005, a theoretical breakthrough showed that some bound entangled states contain secrecy, i.e. can be used for cryptography. In order to explore this topic, I suggest to read first \cite{revqkd}, II.B.2, then the original references given there. However, no explicit protocol to distribute such states has ever been devised, nor will probably ever be: other protocols are much simpler and efficient for the task.

\section{Tutorials}

\subsection{Problems}

\textbf{Exercise 2.1}

Amplification of light is of course compatible with the no-cloning theorem, because spontaneous emission prevents amplification to be perfect \cite{mandel}. Actually, if the amplifier is independent of the polarization, universal symmetric cloning of that degree of freedom is implemented \cite{simon,kempe}. In this problem, we explore the basics of this correspondence.

Consider a single spatial mode of the electromagnetic field and focus on the polarization states; we denote by $\ket{n,m}$ the state in which $n$ photons are polarized $H$ and $m$ photons are polarized $V$. Suppose one photon in mode $H$ is initially present in the amplifier, and that after amplification 2 photons have been produced.
\begin{enumerate}
\item Compute the single-copy fidelity of this cloning process. \textit{Hint:} if you don't remember the physics of amplification, you can reach the result by comparing $a_H^{\dagger}\ket{1,0}$ with $a_V^{\dagger}\ket{1,0}$.
\item How would you describe the state of the system (field + amplifier medium) in this process? \textit{Hint:} compare with the B-H QCM.
\end{enumerate}

\subsection{Solutions}

\textbf{Exercise 2.1}

\begin{enumerate}
\item The theory of spontaneous and stimulated emission implies that, starting with $\ket{1,0}$, the probability of creating $\ket{2,0}$ is twice as large as the probability of creating $\ket{1,1}$. The single-copy fidelity is defined as the probability of finding one of the photons in the initial state, whence obviously $F=\frac{2}{3}\times 1 + \frac{1}{3}\times\frac{1}{2}=\frac{5}{6}$. This is identical to the fidelity for optimal universal symmetric cloning.
\item The analogy with cloning is actually exact: indeed, by conservation of angular momentum, the emission of an $H$ photon and of a $V$ photon cannot be due to the same process. Therefore, after amplification and post-selection of the emission of two photons, the state of the system ``field + amplifying medium'' reads $\sqrt{\frac{2}{3}}\ket{2,0}\otimes\ket{e_H}+\sqrt{\frac{1}{3}}\ket{1,1}\otimes\ket{e_V}$, i.e., in first-quantized notation
\ban
\sqrt{\frac{2}{3}}\ket{H}\ket{H}\otimes\ket{e_H}+\sqrt{\frac{1}{3}}\ket{\Psi^+}\otimes\ket{e_V}
\ean and this exactly the state produced by the B-H QCM.
\end{enumerate}

\chapter{Primitives of quantum information (II)}

\section{State discrimination}
\label{secstatediscr}

Under the head of state discrimination, a large variety of tasks can be accommodated. For a school, rather than reviewing each of them exhaustively, I find it more useful to present concrete examples of each. A review article was written by Chelfes in the year 2000 \cite{chefles}: it contains most of the basic ideas; some recent developments will be mentioned below.

\subsection{Overview}

As the name indicates, state discrimination refers to obtaining information about the quantum state produced by a source that is not fully characterized. We can broadly divide the possible tasks in two categories:
\begin{itemize}
\item \textit{Single-shot tasks:} the goal is to obtain information on each signal emitted by the source. Without further information, as well-known, the task is almost hopeless: basically, after measuring one system, the only thing one can be sure of is that the state was not orthogonal to the one that has been detected. However, the task becomes much more appealing if some additional knowledge is present: for instance, if one is guaranteed that each system can be either in state $\rho_1$ or in state $\rho_2$, with the $\rho_j$ two well-specified states. In such situations, one can consider \textit{probabilistic state discrimination} and try to minimize the probability of a wrong guess; or even \textit{unambiguous state discrimination}, a POVM that either identifies the state perfectly or informs that the discrimination was inconclusive.
\item \textit{Multi-copy tasks:} if the source is guaranteed to produce always the same state, the state can be exactly reconstructed asymptotically; this process is called \textit{state reconstruction}, or \textit{state estimation}, or \textit{state tomography}. If, in addition, one knows that the state is either $\rho_1$ or $\rho_2$, one can ask how fast the probability of wrong guess decreases with the number of copies and obtain a quantum version of the \textit{Chernoff bound}. If the source is not guaranteed to produce always the same state, the task seems hopeless, and in full generality it is; but if the observed statistics are symmetric under permutation, a quantum version of the \textit{de Finetti theorem} exists.
\end{itemize}

\subsection{Single-shot tasks (I): Probabilistic discrimination}
\label{ssprobadiscr}

\subsubsection{Probabilistic discrimination of two states}

Let us consider the simplest case: two states $\rho_1$ and $\rho_2$ are given each with probability $\eta_1$ and $\eta_2=1-\eta_1$. At each run, one performs a measurement, whose outcome is used to guess which state was given; we want to \textit{minimize the probability $P_{\mathrm{error}}$ that the guess is wrong}.

Since ultimately we want two outcomes, without loss of generality the measurement can be described by two projectors $\Pi_1$ and $\Pi_2=\one-\Pi_1$ where $\one$ is the identity over the subspace spanned by $\rho_1$ and $\rho_2$. Therefore, the probability of guessing $\rho_j$ correctly, i.e. the probability of guessing $j$ given $\rho_j$, is given by $\Tr(\Pi_j\rho_j)$; whence the average probability of error for this measurement is
\ba
P_{\mathrm{error}}(\Pi_1)&=&1-\sum_{j=1,2}\eta_j \Tr(\Pi_j\rho_j)\,=\,\eta_1-\Tr\big(\Pi_1(\eta_1\rho_1-\eta_2\rho_2)\big)\,.
\ea
In order to minimize this, we have to find the projector $\Pi_1$ that maximizes the second term on the right-hand side. The result is \cite{helstrom,herzog}
\ba P_{\mathrm{error}}&=&\demi\big[1-\Tr\left|\eta_1\rho_1-\eta_2\rho_2\right|\big]\,.\label{perrorgen}\ea A constructive measurement strategy that would lead to this optimal result is the following: measure the Hermitian operator $M=\eta_1\rho_1-\eta_2\rho_2$; if the outcome is a positive eigenvalue, guess $\rho_1$, if it's a negative eigenvalue, guess $\rho_2$.

Let us prove (\ref{perrorgen}) for the special case of \textit{equal a priori probabilities} $\eta_1=\eta_2=\demi$. In this case, $\Pi_1$ is the projector on the subspace of positive eigenvalues of $\rho_1-\rho_2$; but since $\Tr(\rho_1-\rho_2)=0$, the sum of the positive eigenvalues and of the negative ones must be the same in absolute value. Therefore we find $\max_{\Pi_1}\Tr\big(\Pi_1(\rho_1-\rho_2)\big)=\demi\Tr\left|\rho_1-\rho_2\right|$ and finally \ba P_{\mathrm{error}}=\demi\left[1-\demi\Tr\left|\rho_1-\rho_2\right|\right]&\;&\mbox{ $(\eta_1=\eta_2=\demi)$.}\label{perror}\ea

\subsubsection{Intermezzo: trace distance}

The mathematical object
\ba
D(\rho,\sigma)&=&\demi\,\Tr \big|\rho-\sigma\big|
\ea
that appeared in the previous proof is called \textit{trace-distance} between two states $\rho$ and $\sigma$. This quantity appears often in quantum information, so it is worth while spending some time on it (for all proofs and more information, refer to chapter 9 of \cite{nielsen}). Some simple properties of the trace distance are: $D(\rho,\sigma)=0$ if and only if $\rho=\sigma$; the maximal value $D(\rho,\sigma)=1$ is reached for orthogonal states; $D(\rho,\sigma)=D(\sigma,\rho)$. It can moreover be proved that the triangle inequality $D(\rho,\tau)\leq D(\rho,\sigma)+D(\sigma,\tau)$; therefore it has the mathematical properties of a ``metric'', i.e. it defines a valid distance between states.

As we have seen, two states can be distinguished with probability at most $\demi[1+D]$. But $\demi$ is sheer random guessing: rewriting $\demi[1+D]=D\times 1+(1-D)\times\demi$, we see that, \textit{in any task}, the two states behave differently with probability at most $D$. Indeed, suppose there is a task for which the two behaviors are more distinguishable: we would use that task as measurement for discrimination, thus violating the bound (\ref{perror}).

The usefulness of this remark becomes even more apparent when phrased in a slightly different context. Instead of having to discriminate two states, suppose one has a state $\rho$ and wants to compare it to an ``ideal'' state $\rho_{\mathrm{ideal}}$. Then $D(\rho,\rho_{\mathrm{ideal}})$ is the \textit{maximal probability of failure}, i.e. the maximal probability that the real state will produce a result different than the one the ideal state would have produced. This plays a central role, for instance, in the definition of security in quantum cryptography; see II.C.2 in \cite{revqkd} for a discussion and original references. We shall find this idea in a different context in Section \ref{devindeptomo}.

\subsubsection{PSD of more than two states}

In the most general case, as expected, the optimal PSD strategy is not known. Sub-optimal strategies are trivially found: just invent a measurement strategy and compute the probabilities of failure. A particular strategy performs often quite well, so much so that it has been called \textit{pretty good measurement (PGM)} \cite{pgm}. It is defined as follows: let $\{\rho_k,p_k\}$ be the set of states to be distinguished, with the corresponding a priori probabilities; and define $M=\sum_k p_k\rho_k$. Then the PGM is the POVM whose elements are defined by $E_k=p_k M^{-1/2}\rho_k M^{-1/2}$. The special case, in which all the states to be distinguished are pure ($\rho_k=\ket{\psi_k}\bra{\psi_k}$) and equally probable, is known as \textit{square-root measurement}; in this case, the elements of the POVM are the suitably normalized projectors on the states $\ket{\chi_k}\propto M^{-1/2}\ket{\psi_k}$ \cite{haus2}. There is a significant amount of literature on these measurements, that in some case define the optimal measurement strategy. Its review goes beyond our scope.

Let us finally mention that an \textit{upper bound} on the guessing probabilities can be found by studying the dual problem\footnote{The dual problem is related to the definition of min-entropies. Let $\{(\rho_k,\eta_k)\}_{k=1...n}$ be the states to be distinguished with the respective a priori probabilities. Then one forms the \textit{classical-quantum} state $\rho_{AB}=\sum_k\eta_k \ket{k}\bra{k}\otimes\rho_k$. Choose now a state $\sigma_B$ and compute the minimum $\lambda$ such that $M=\lambda\one_A\otimes\sigma_B-\rho_{AB}$ is a non-negative operator: then $\lambda\geq 1-P_{\mathrm{error}}$, with the guarantee of equality for the minimum over all possible $\sigma_B$.} \cite{dualhmin}.

\subsubsection{PSD and cloning}

There is an interesting, somehow intuitive link between optimal cloning and PSD, namely: for any ensemble of \textit{pure} states $\{(\ket{\psi_k},\eta_k)\}_{k=1...n}$ to be distinguished, optimal PSD is equivalent to optimal symmetric $1\rightarrow N$ cloning in the limit $N\rightarrow \infty$ \cite{bae}. Note that this fact does not help to find the explicit strategy, since optimal state-dependent cloners are not known in general either.

The argument goes as follows. For convenience, define $F_C$ as the single-copy fidelity of the optimal $1\rightarrow\infty$ cloner and $F_M$ as the fidelity of the state reconstructed after the optimal PSD measurement. It is obvious that $F_M\leq F_C$: after measurement, we have a guess for the state, so we can just create as many copies as we want of that state, so this defines a possible cloner. The proof of the converse is more tricky. Basically, one applies the cloner $C$ to a half of a maximally entangled state: $(\one\otimes C)\ket{\Phi^+}=\rho_{AB_1...B_N}$. By assumption, all the $\rho_{AB_j}$ are equal. In the limit $N\rightarrow\infty$ the information of $B_j$ is ``infinitely shareable'' and a theorem then guarantees that $\rho_{AB_j}$ is separable. But then, the restriction $\tilde{C}$ defined as $(\one\otimes \tilde{C})\ket{\Phi^+}=\rho_{AB_1}$ describes an entanglement-breaking channel, and it is known that any such channel is equivalent to performing a measurement and forwarding the collapsed state. In conclusion, there exist a measurement strategy that achieves the single-copy fidelity of the optimal cloner.

\subsection{Single-shot tasks (II): Unambiguous state discrimination}

We consider now a different situation: now we want discrimination to be unambiguous; the price to pay is that sometimes the procedure will output an inconclusive outcome.

\subsubsection{Unambiguous discrimination of two pure states}

The case where the two state have equal a priori probabilities was solved independently by Ivanovics, Dieks and Peres \cite{iva,dieks88,peres88}; Jaeger and Shimony later solved the problem for arbitrary probabilities \cite{jaegershimony}. See also \cite{peres}, sect. 9-5. For two \textit{pure} states, the discrimination succeeds at most with probability $1-|\braket{\psi_1}{\psi_2}|$; of course, with probability $|\braket{\psi_1}{\psi_2}|$ one obtains the inconclusive outcome.

\begin{center}
\begin{figure}
\includegraphics[scale=0.8]{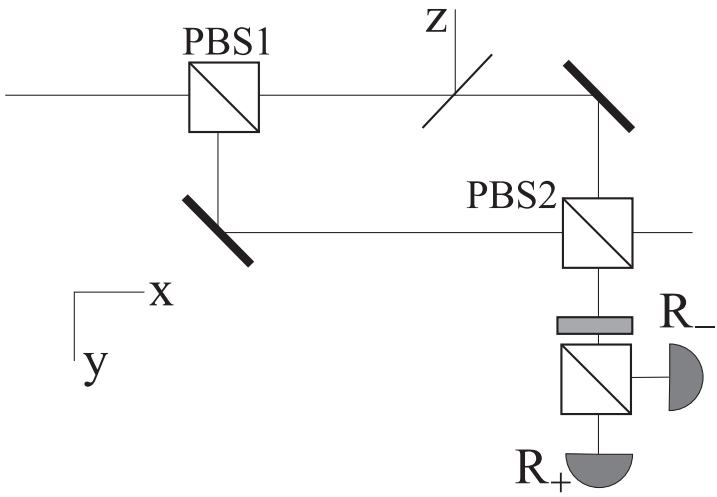}
\caption{Optical setup for unambiguous discrimination of non-orthogonal polarization states.}\label{figunamb}
\end{figure}
\end{center}

Let us begin by describing a simple setup that achieves USD of two pure states \cite{huttner}; an even simpler one is given as a tutorial. The setup is shown in figure \ref{figunamb}. The polarizing beam-splitters PBS1 and PBS2 are oriented such that $\ket{H}$ is transmitted and $\ket{V}$ is reflected. The goal is to distinguish between
\ba
\ket{\psi_{\pm}}&=&\cos\alpha\ket{H}\pm\sin\alpha\ket{V}
\ea
whose overlap is $\chi=\cos 2\alpha$. Indeed, we have: \ban
\big(\cos\alpha\ket{H}\pm\sin\alpha\ket{V}\big)\ket{x}&
\stackrel{PBS1}{\rightarrow}&
\cos\alpha\ket{H}\ket{x}\,\pm\,\sin\alpha\ket{V}\ket{y}\\
&\stackrel{BS\,x-z}{\rightarrow}
&\cos\alpha\sqrt{t}\,\ket{H}\ket{x}\,\pm\,\sin\alpha\ket{V}\ket{y}
\,+\, \cos\alpha\sqrt{1-t}\,\ket{H}\ket{z}\\
&\stackrel{mirrors}{\rightarrow}&
i\cos\alpha\sqrt{t}\,\ket{H}\ket{y}\,\pm\,i\sin\alpha\ket{V}\ket{x}
\,+\, \cos\alpha\sqrt{1-t}\,\ket{H}\ket{z} \\
&\stackrel{PBS2}{\rightarrow}&
i\big(\cos\alpha\sqrt{t}\,\ket{H}\,\pm\, \sin\alpha\ket{V}\big)
\ket{y} \,+\, \cos\alpha\sqrt{1-t}\,\ket{H}\ket{z}\,. \ean If $\cos\alpha\sqrt{t}=\sin\alpha$, that is for transmission of the BS $t=\tan^2\alpha$, then the two states that can appear in mode $y$ are orthogonal and can be discriminated by a usual projective measurement (output modes $\ket{R_{\pm}}$ of a PBS). In summary, we have implemented the transformation \ba
\big(\cos\alpha\ket{H}\pm\sin\alpha\ket{V}\big)\ket{x}&
\longrightarrow&
  i\sqrt{2}\sin\alpha\,\ket{\pm}\ket{R_{\pm}}\,+\,\cos\alpha\sqrt{1-\tan\alpha}\,
  \ket{H}\ket{z}
\ea where the probability of obtaining a conclusive result is the
optimal one, since $2\sin^2\alpha=1-\cos 2\alpha=1-|\braket{\psi_+}{\psi_-}|$.

Taking a more formal view, this setup realizes a POVM: there are three outcomes for a qubit. The two conclusive outcomes are described by
\ba
A_{+}=\eta\ket{+}\bra{\ket{\psi_-^{\perp}}}&,& A_{-}=\eta\ket{-}\bra{\ket{\psi_+^{\perp}}}
\ea
with $\ket{\psi_{\pm}^{\perp}}=\sin\alpha\ket{H}\mp\cos\alpha\ket{V}$. The factor $\eta$ is determined by the constraint that the largest eigenvalue of $A_{+}^\dagger A_{+}+A_{-}^\dagger A_{-}$ should be 1; direct inspection leads to $\eta=\frac{1}{\sqrt{2}\,\cos\alpha}$. The probability of guessing correctly $\ket{\psi_\omega}$ is given by $\Tr\left(A_{\omega}\ket{\psi_\omega}\bra{\psi_\omega}A_{\omega}^\dagger\right)=2\sin^2\alpha=1-\cos 2\alpha$, as it should.

\subsubsection{Generalizations: more pure states, mixed states}

For any number \textit{pure states}, unambiguous state discrimination is possible if and only if the states are linearly independent. This implies in particular that no set consisting of $N>d$ states, where $d$ is the dimension of the Hilbert space, can be discriminated unambiguously.

Though not optimal in general, the following strategy can always be implemented. Consider $N$ linearly independent states $\{\ket{\psi_k}\}_{k=1,...,N}$ generating the $N$-dimensional subspace ${\mathcal E}$. Let $\ket{\varphi_k}$ be the vector in ${\mathcal E}$ that is orthogonal to all the vectors except $\ket{\psi_k}$: of course, if one performs a measurement and finds the result $\ket{\varphi_k}$, one can unambiguously identify $\ket{\psi_k}$ as being the measured state. This looks very much like a usual projective measurement defined by the $P_k=\ket{\varphi_k}\bra{\varphi_k}$; however, the $\ket{\varphi_k}$'s are in general not orthogonal (if they are all orthogonal, they coincide with $\ket{\psi_k}$ that are therefore already orthogonal), so $\sum_kP_k$ does not need to be proportional to $\one$ and its largest eigenvalue may be $\lambda>1$. A possible way of obtaining a valid POVM consists in defining $A_k=P_k/\sqrt{\lambda}$ and associate the inconclusive outcome to $A^\dagger_0A_0=\one-\sum_kP_k/\lambda$.

When it comes to unambiguous discrimination of \textit{mixed states}, the situation is far more complex. For instance, even two different mixed states cannot be discriminated unambiguously if their supports are identical (obviously, two pure states have identical support if and only if they are the same state). Some non-trivial examples have been worked out \cite{raynal}.

\subsection{Multi-copy tasks (I): tomography, Chernoff bound}

\subsubsection{Tomography of an a priori completely unknown state}

The idea of tomography is inherent to the notion of state itself. A state is a description of our statistical knowledge. In classical statistics, a probability distribution can be reconstructed by collecting many independent samples that follow the distribution. Similarly, if one performs suitable measurements on $N$ copies of an unknown quantum state $\rho$, one can obtain a faithful description of the state itself.

Some examples will suffice to clarify the principle. Any one-qubit state can be written as $\rho=\demi\left(\one+\vec{m}\cdot\vec{\sigma}\right)$ where $m_k=\moy{\sigma_k}$. Therefore, if one can estimate the average values $\moy{\sigma_x}$, $\moy{\sigma_y}$ and $\moy{\sigma_z}$, one can reconstruct the state. Similarly, for two qubits, one has to estimate the fifteen average values $\moy{\sigma_k\otimes\one}$ (three), $\moy{\one\otimes\sigma_k}$ (three) and $\moy{\sigma_j\otimes\sigma_k}$ (nine). It's exactly like reconstructing a classical distribution, but for the fact that one has to perform several different samplings (measurements) because of the existence of incompatible physical quantities.

While the principle is simple and necessary for the notion of state to have any meaning, the topic is not closed. For instance, one can look for ``optimal'' tomography according to different figure of merit (smallest size of the POVM, faster convergence with the number of samples $N$, etc.).

Among the rigorous results that have been obtained only recently, let us mention the computation of the \textit{quantum Chernoff bound}. The task is somehow half-way between state estimation and tomography: given the promise that the state is either $\rho_1$ or $\rho_2$, one wants to estimate \textit{how fast} the probability of distinguishing increases for increasing $N$. The result is most easily formulated in terms of the probability of error, which has been proved to decrease exponentially: $\demi\left[1-\demi\left|\eta_1\rho_1^{\otimes N}-\eta_2\rho_2^{\otimes N}\right|\right]\sim e^{-\xi N}$ with $\xi=-\log\left(\min_{0\leq s\leq 1}\Tr(\rho_0^s\rho_1^{1-s})\right)$ \cite{chernoff}.

\subsection{Multi-copy tasks (II): de Finetti theorem and extensions}

The tomography discussed above works only under the assumption that the source produces always the same state, i.e. that the state of $N$ emitted particles is simply $\rho^{\otimes N}$. Operationally, this cannot be guaranteed: all that one can guarantee is that the experimental procedure is the same for each realization. This implies that the $N$-particle state $\xi_N$, whatever it is, is such that all possible statistics are invariant by permutation of the particles (for we have no information that allows to distinguish some realizations from others).

Now, the statement we want to make is the following: \textit{in the limit of large $N$, any state $\xi_N$ that is invariant by permutation is ``close'' to a product state $\rho^{\otimes N}$} or to a classical mixture of such states\footnote{The careful reader may immediately notice that the statement, as loosely stated, cannot be strictly true: for instance, the GHZ state $\ket{000...0}+\ket{111...1}$ is invariant under permutation but is far, under any measure, from a product state. Indeed, the exact statement says that: if $\xi_N$ is invariant under permutation, there exist $k<N$ such that $\Tr_k\xi_N$ is close to a mixture of product states (see how this is the case for the GHZ state, already with $k=1$).}. It is not superfluous to notice that we use this result, implicitly, in every laboratory experiment, be it about classical or quantum physics.

There are several proofs of this statement, differing in important details (what ``to be close'' exactly means) and consequences. Most of them are called ``de Finetti theorems'', from the name of the Italian mathematician who first proved a similar statement in the context of classical probability theory; they consist on an estimate of the trace distance between the real state and the set of product states. For a very clear presentation, including a review of previous works, we refer the student to \cite{rennernphys}. The most recent extension, that obtains a much better bound under relaxed assumptions, is based on the idea of post-selection \cite{rennerpostsel}.

\section{Quantum coding}

By \textit{quantum coding} I denote the generalization to quantum physics of the main results of classical information theory. A very thorough presentation of the basic material is already available in chapters 11 and 12 of Nielsen and Chuang's book \cite{nielsen}: unless other references are given, this section is a distillation of that source, to which the reader should also refer to obtain the references to the original works. The studies of security of quantum cryptography are of course an offspring of this field, so several notions will be presented in due detail in the specific series of lectures.

\subsection{Shannon entropy and derived notions}

It should be well-known that ``entropy'' is associated first with ``uncertainty'' in information theory. The elementary entropic quantity in quantum physics is the \textit{von Neumann entropy}, which is nothing else than the Shannon entropy of the eigenvalues $\{\lambda_k\}$ of a given state $\rho$:
\ba
S(\rho)&=&-\Tr(\rho\log\rho)\,=\,-\sum_{k}\lambda_k\log\lambda_k\,.
\ea
Many of the properties of Shannon entropy translate directly to von Neumann entropy. For instance, it's a concave function: the entropy of a mixture is larger than the average of the entropies, formally
\ba
S\Big(\sum_kp_k\rho_k\Big)&\geq& \sum_kp_k S\left(\rho_k\right)\,.
\ea
One can also define \textit{relative entropy} $S(\rho||\sigma)=-S(\rho)-\Tr(\rho\log\sigma)$, with similar properties as the classical analog. Of course, the quantum \textit{joint entropy} is just defined as $S(A,B)=S(\rho_{AB})$. Again, it behaves like the classical analog; in particular, the property called ``strong subadditivity'' holds: $S(A,B,C)+S(B)\leq S(A,B)+S(B,C)$.

Not exactly everything is a copy of classical information theory, though: a most remarkable exception is the possibility for \textit{conditional entropy} to be negative. Classical conditional entropy for a probability distribution $P(a,b)$ is the uncertainty on the distribution on $A$ knowing $B$: it's the entropy of the conditional probabilities, averaged over all possible conditions, i.e. $H(A|B)=\sum_{b}P(b)\Big[-\sum_{a}P(a|b)\log P(a|b)\Big]$. This definition cannot be generalized as such in quantum physics, because there is no obvious analog of $P(a|b)$; however, it is simple to show that $H(A|B)=H(A,B)-H(B)$, and this expression can be generalized:
\ba
S(A|B)&=&S(\rho_{AB})-S(\rho_B)\,.
\ea
Consider now, as an example, a maximally entangled state of two qubits $\rho_{AB}=\ket{\Phi^+}\bra{\Phi^+}$: this state is pure, so $S(\rho_{AB})=0$; however, $\rho_B$ is maximally mixed, whence $S(\rho_B)=1$; so in all $S(A|B)=-1$. This is a manifestation of one of the unexpected features of entanglement: the fact that one has sharp properties for the composite system that do not derive from sharp properties of the sub-systems.

Interestingly, conditional entropy has an operational interpretation as the amount of quantum information needed to transfer a state $\rho_{AB}$, initially shared between two parties, to one of the parties, while keeping the coherence with possible purifying systems that are not available to either party \cite{how05}. When the quantity is negative, it basically means that, after the transfer, the parties still keep some quantum correlations\footnote{In the extreme case of pure states, this is pretty clear. Alice and Bob know which state they share. If the state is pure, there is no coherence with a purifying system to be preserved! Therefore, Bob can just generate the state in his own location: no resources are used, and the shared states are still available for teleportation.} that can be used to transfer additional quantum informations via teleportation. 

\subsection{From Holevo to Schumacher, and beyond}

We start by presenting the two best-known early results in quantum coding: the Holevo bound and Schumacher's compression. 

\subsubsection{Holevo bound: classical information coded in quantum states}

Suppose Alice prepares any of the states $\{\rho_x\}_{x=1...N}$, each with some probability $p_x$ and sends them to Bob on a noiseless channel. Bob performs a measurement, generically a POVM, with outcomes $y\in\{1,..., m\}$. Since ultimately both Alice's and Bob's variables are classical, the amount of information that Bob has obtained through the measurement is given by the Shannon mutual information $I(X:Y)=H(X)+H(Y)-H(X,Y)$. The \textit{Holevo bound} is an upper bound on this amount of information:
\ba
I(X:Y)&\leq& \chi_{\{p_x,\rho_x\}}\,\equiv\,S\big(\sum_xp_x\rho_x\big)-\sum_xp_xS(\rho_x)\,.
\ea
The left-hand side is a purely classical quantity, because $X$ and $Y$ are classical; the right-hand side is quantum, because classical information has been coded in quantum states. The bound can be saturated only if the states $\rho_x$ are mutually commutative. We shall meet below an extension of this bound.

Maybe it's convenient to stress something at this point. The configuration envisaged in this paragraph looks at first like a mathematical exercise with little or no usefulness in practice. Indeed, the task is ``Alice wants to send a non-secret message to Bob''. To achieve this, Alice most probably would not use quantum states at all in the first place; or more precisely, she'd use orthogonal states (this is classical communication), so that Bob would have no trouble discriminating them. So, why should one bother about such an artificial task as sending messages with non-orthogonal states? The answer is the following: in itself, the situation is artificial indeed; but this situation may appear as natural in the context of another task. For instance, consider the effective channel linking Alice to Eve in quantum cryptography. Here, Alice does \textit{not} want Eve to learn the message: it's Eve that sneaks in and get whatever she can. It is therefore normal that the channel Alice-Eve is not optimized for direct communication. It turns out that this channel is precisely of the Holevo type: Alice's bits are encoded in quantum states that Eve keeps (see also Tutorial, and the series of lectures on quantum cryptography in this school).

\subsubsection{Schumacher compression of quantum information}

Consider a source producing classical symbols $x$, independent and identically distributed (i.i.d.) according to a distribution $p(x)$. One wants to know how much a message can be compressed and later decompressed without introducing errors, i.e. $m(n)$ in $(x_1,...,x_n)\stackrel{{\mathcal C}}{\longrightarrow} (y_1,...,y_m) \stackrel{{\mathcal D}}{\longrightarrow}(x_1,...,x_n)$. A well-known theorem by Shannon says that, asymptotically, $m(n)=nH(X)$.

Schumacher's coding is the quantum generalization of this theorem. The source is now represented by the mixed state $\rho=\sum_x p(x)\ket{x}\bra{x}$ living in a $d$-dimensional Hilbert space ${\mathcal H}$ (i.e. $\log d$ qubits). The compression-decompression procedure is now $\rho^{\otimes n}\stackrel{{\mathcal C}}{\longrightarrow} \sigma \stackrel{{\mathcal D}}{\longrightarrow}\rho^{\otimes n}$. The result is that $\sigma$ must live in a Hilbert space of dimension $2^{nS(\rho)}$.  

\subsubsection{The richness of quantum information theory}

With classical signals, one can basically send classical information. When one realizes the huge body of knowledge that this has generated \cite{cover}, the complexity of opening the box to quantum physics becomes striking. Indeed, once quantum systems are brought into the game, the number of possible situations that one can envisage explodes:
\begin{itemize}
\item The \textit{nature of the coding}: classical information theory has classical bits (\textit{c-bits}). In quantum information theory, one has of course to add quantum bits (\textit{q-bits or qubits}), but also more complex units like bits of entanglement (\textit{e-bits}) because states of two qubits are not the same of two states of qubits.
\item The \textit{available resources}: in classical information theory, one has classical channels for communication and shared randomness for possible pre-established correlations. Now we have to add quantum channels (those that allow to send qubits), pre-established entanglement... For instance, one can study what can be done with a classical channel assisted with shared entangled pairs; and all possible combinations.
\end{itemize}

Many results we have reviewed, including Schumacher's, may give the impression that ultimately one will always find the analog of known results in classical information theory. However, this is \textit{not} the case, and the reason is \textit{entanglement}. Indeed, a crucial assumption in Schumacher's result is that the source is i.i.d. We are going to gain more insight on this point by studying \textit{channel capacities}, a field in which several ground-breaking results have been found only recently.

\subsection{Channel capacities: a rapid overview}

\subsubsection{Definition of quantum channel capacity}

Alice wants to send $m$ x-bits to Bob (x can be c, q, e...). She encodes her information in a state of $n$ qubits (possibly entangled) and sends these qubits over a quantum channel $T$ to Bob, who decodes the information correctly with probability $\varepsilon(m,n)$. The \textit{capacity} of this channel is given by
\ba
C_x(T)=\sup\,\lim_{n\rightarrow\infty}\frac{m}{n}&\mbox{ such that }&\lim_{n\rightarrow\infty}\varepsilon(m,n)\,=\,0\,.
\ea
The supremum is taken over all possible choice of coding and decoding. In general therefore, for a given quantum channel $T$, several different capacities can be defined, depending on the nature of the information to be transmitted (x-bits), but also on possible additional resources, on the requested speed of convergence of the error rate etc.  

\subsubsection{Case study: classical capacity of a quantum channel}

For definiteness, let us focus on the \textit{classical capacity} of a quantum channel, written ${\mathcal C}(T)$. The scenario is the one considered by Holevo: Alice codes classical information $X$ in quantum states, sends them to Bob who performs a POVM and extracts classical information Y. The only difference is that now the quantum systems are sent over a non-trivial channel $T$: i.e., if Alice sends $\rho_k$, Bob receives $T[\rho_x]$.

Under the assumption of an i.i.d. source, we can easily understand that
\ba
{\mathcal C}_{i.i.d.}(T)&=& \chi(T)\,=\,\max_{p_x,\rho_x}\,\chi_{\{p_x,T[\rho_x]\}}\,.\label{ciid}
\ea
But is it possible to achieve a larger capacity with a non i.i.d. source? In classical information, this is known not to help: one can always maximize the rate of a channel with i.i.d. sources. In this case, one says that \textit{capacity is additive}. But quantum physics allows for entanglement: Alice may associate the sequence $(x_1,x_2)$ to an entangled state $\rho_{x_1x_2}$. Can this help? For long time, the answer was conjectured to be negative, though explicit proofs were available only for some particular channels. In September 2008, however, Hastings proved that the conjecture is in general wrong: there exist channels, whose full classical capacity cannot be reached by i.i.d. sources --- in other words, for some channels, entanglement does help even if you are using the channel to share classical information!

Because of this, the general expression of the classical capacity of a quantum channel is
\ba
{\mathcal C}(T)\,=\,\lim_{n\rightarrow\infty}\frac{1}{n}\chi(T^{\otimes n})&\mbox{ with } &\chi(T^{\otimes n})\,=\,\max_{p_\mathbf{x},\rho_\mathbf{x}}\,\chi_{\{p_\mathbf{x},T^{\otimes n}[\rho_\mathbf{x}]\}}\,:
\ea
now the maximum must be taken over all possible choice of $n$-qubit states, each coding for the classical information $\mathbf{x}=(x_1,...,x_n)$. There is manifestly no hope of computing such a maximum by brute force. This is the reason why even the ``simple'' classical capacity of quantum channels in not known for arbitrary channels.

\subsubsection{Other capacities}

Among the other possible capacities of a quantum channel, two are worth at least mentioning:
\begin{itemize}
\item The \textit{classical private capacity} is associated to the task of sending classical information while keeping it secret from the environment. Its expression is \cite{devetak02}:
\ba
{\mathcal P}(T)&=& \lim_{n\rightarrow\infty}\frac{1}{n} \max_{p_\mathbf{x},\rho_\mathbf{x}}\,\left[\chi_{\{p_\mathbf{x},T^{\otimes n}[\rho_\mathbf{x}]\}}\,-\, \chi_{\{p_\mathbf{x},\tilde{T}^{\otimes n}[\rho_\mathbf{x}]\}}\right]\,.
\ea
where $\tilde{T}$ is the ``complementary channel'', i.e. the information that leaks into the environment --- as such, the formula is somehow intuitive.
\item The \textit{quantum capacity} is the capacity of the channel when Alice wants to send quantum information. It is generally written ${\mathcal Q}$; its expression would require introducing additional notions that are beyond the scope of our survey.
\end{itemize}
In general, it holds
\ba
{\mathcal C}(T)&\geq\,{\mathcal P}(T)&\geq\, {\mathcal Q}(T)\,.
\ea
The fact that the first inequality can be strict is almost obvious; less obvious is the fact that the second inequality can also be strict \cite{horodopp}. The proofs of non-additivity are all very recent: Smith and Yard had proved the non-additivity of the quantum capacity in Summer 2008 \cite{sy08}. Later, the conjecture that the private capacity may also be non-additive was formulated \cite{privatecap} and proved \cite{liwinter}. At the moment of writing, the field is still very active.

\section{Tutorials}

\subsection{Problems}

\noindent\textbf{Exercise 3.1}

Prove that the trace distance between two pure states $\ket{\psi_1}$ and $\ket{\psi_2}$ is given by
\ba D(\rho_1,\rho_2)&=&\sqrt{1-|\braket{\psi_1}{\psi_2}|^2}\,.
\ea \textit{Hint:} Note that you can always find a basis in which $\ket{\psi_1}=c\ket{0}+s\ket{1}$ and $\ket{\psi_2}=e^{i\varphi}(c\ket{0}-s\ket{1})$ with $c=\cos\theta$ and $s=\sin\theta$.\\

\noindent\textbf{Exercise 3.2}

A short laser pulse can be sent either at time $t_1$ or at time $t_2$. By detecting the time of arrival, one can obviously discriminate between these two cases. This rather trivial process is actually an example of \textit{unambiguous state discrimination} of the two two-mode coherent states $\ket{\psi_1}=\ket{0}\ket{\alpha}$ and $\ket{\psi_2}=\ket{\alpha}\ket{0}$; it is used to create the raw key in the quantum cryptography protocol called COW \cite{cow}. We recall the decomposition of the coherent state $\ket{\alpha}$, $\alpha\in\compl$, on the number basis:
\ba
\ket{\alpha}&=& e^{-|\alpha|^2/2}\,\sum_{n=0}^{\infty}\frac{\alpha^n}{\sqrt{n!}}\,\ket{n}\,.\label{coherent}\ea
\begin{enumerate}
\item What is the probability of success for optimal USD?
\item Prove that the POVM for optimal USD can simply be realized by detecting the time of arrival (with a perfect detector). \textit{Hint:} What are the ``inconclusive'' events?
\item Discuss what happens if the detector is not perfect, in particular how the discussion is modified by (i) efficiency $\eta<1$; (ii) dark counts. 
\end{enumerate}

\noindent\textbf{Exercise 3.3}

Let $\{\ket{e_k}\}_{k=1...4}$ be an orthonormal set of four vectors. We define $\ket{\psi_1^{\pm}}=\sqrt{1-\varepsilon}\ket{e_1}\pm \sqrt{\varepsilon}\ket{e_2}$ and $\ket{\psi_2^{\pm}}=\sqrt{1-\varepsilon}\ket{e_3}\pm \sqrt{\varepsilon}\ket{e_4}$; and we construct in turn the mixtures
$\rho_0=(1-\varepsilon)\ket{\psi_1^{+}}\bra{\psi_1^{+}}+\varepsilon\ket{\psi_2^{+}}\bra{\psi_2^{+}}$ and $\rho_1=(1-\varepsilon)\ket{\psi_1^{-}}\bra{\psi_1^{-}}+\varepsilon\ket{\psi_2^{-}}\bra{\psi_2^{-}}$.
\begin{enumerate}
\item Compute the Holevo bound $\chi(\rho_0,\rho_1)$, assuming $p_0=p_1=\demi$.
\item The states given above, of course, have a meaning: they describe Eve's states in the optimal eavesdropping on the BB84 protocol of quantum cryptography, when an error rate $\varepsilon$ is measured by Alice and Bob --- see other series of lectures; and paragraph III.B.2 and Appendix A of \cite{revqkd}. In this scenario, what does the Holevo bound mean? \textit{Hint:} the index $a$ or the matrices $\rho_a$ is Alice's bit.
\end{enumerate}

\subsection{Solutions}

\noindent\textbf{Exercise 3.1}

By writing $\ket{\psi_1}=c\ket{0}+s\ket{1}$ and $\ket{\psi_2}=e^{i\varphi}(c\ket{0}-s\ket{1})$, we have $\braket{\psi_1}{\psi_2}=e^{i\varphi}(c^2-s^2)=e^{i\varphi}\cos 2\theta$ and
\ban
\rho_1-\rho_2&=&2cs\sigma_x
\ean
whence $D(\rho_1,\rho_2)=\demi(|+2cs|+|-2cs|)=2cs=\sin 2\theta$. The result follows immediately.\\

\noindent\textbf{Exercise 3.2}
\begin{enumerate}
\item The probability of optimal USD is $p_{USD}=1-|\braket{\psi_1}{\psi_2}|$; here $|\braket{\psi_1}{\psi_2}|=|\braket{0}{\alpha}|^2=e^{-|\alpha|^2}$.
\item As soon as the detector fires, the two states can be distinguished; so the ``inconclusive'' events are the events in which the detector did not fire; if the detector has perfect efficiency, this can only happen because of the vacuum component of the state. In both $\ket{\psi_1}$ and $\ket{\psi_2}$, the amplitude of the vacuum component is $e^{-|\alpha|^2/2}$; therefore the probability that the detector fires is $1-e^{-|\alpha|^2}=p_{USD}$.
\item A detector with efficiency $\eta<1$ is equivalent to losses $\sqrt{\eta}$; the discrimination is still unambiguous but succeeds only with probability $p=1-e^{-\eta|\alpha|^2}<p_{USD}$ (note that this is still the optimal procedure, under the constraint that one has to use such imperfect detectors). If dark counts are present, the detector may fire even if there was no photon; therefore the discrimination is no longer unambiguous.
\end{enumerate}

\noindent\textbf{Exercise 3.3}
\begin{enumerate}
\item We have to compute $\chi(\rho_0,\rho_1)=S(\rho)-\demi[S(\rho_0)+S(\rho_1)]$ with $\rho=\demi(\rho_0+\rho_1)$. Now, $\rho_0$ and $\rho_1$ are both incoherent mixtures of two orthogonal states with the same weights; therefore $S(\rho_0)=S(\rho_1)=-(1-\varepsilon)\log(1-\varepsilon)-\varepsilon\log\varepsilon\equiv h(\varepsilon)$. Moreover, $\rho=(1-\varepsilon)^2\ket{e_1}\bra{e_1}+\varepsilon(1-\varepsilon)\ket{e_2}\bra{e_2}+\varepsilon(1-\varepsilon) \ket{e_3}\bra{e_3}+\varepsilon^2\ket{e_4}\bra{e_4}$, whence $S(\rho)=2h(\varepsilon)$. All in all, $\chi(\rho_0,\rho_1)=h(\varepsilon)$.
\item One can see the relation between Alice and Eve as a channel, in which Alice's bit value $a$ has been encoded in a state $\rho_a$. Therefore, the Holevo bound represents the maximal information that Eve might extract about Alice's bit.
\end{enumerate}

\chapter{Quantum correlations (I): the failure of alternative descriptions}

This and the following two lectures are devoted to the violation of Bell's inequalities and related topics. This means that we shall focus on a restricted family of quantum phenomena: the establishment of \textit{correlations between distant partners through separated measurement of entangled particles}. This kind of phenomenon by no means exhausts the possibilities of quantum physics. However, it is there that the discrepancy between classical and quantum physics manifests itself in the most straightforward way (this Lecture). As such, one might expect this feature to be useful for some quantum information tasks: this is indeed the case, but curiously enough, this awareness is only rather recent (Lecture 6). Lecture 5 will be an excursion into an extended theoretical framework that may be very promising or may just be a wrong track, but is funny and worth at least having heard about.

\section{Correlations at a distance}

\subsection{Generic setup under study}

\begin{figure}[ht]
\includegraphics[scale=0.9]{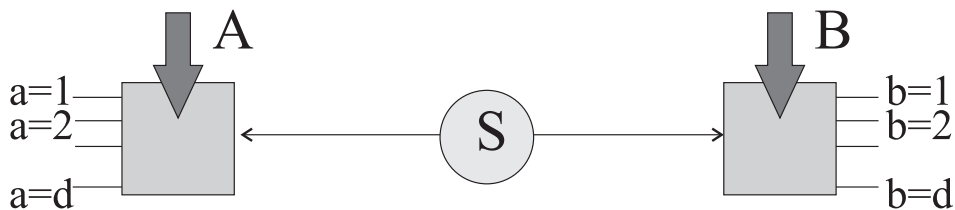}
\caption{The setup to be kept in mind for Bell-type experiments: a source S distributed two quantum systems to separated locations. On each location, the physicist is free to choose which measurement to perform ($A$, $B$); as a result, they obtain outcomes $a$, $b$. Of course, this setup can be generalized to more parties, or to the case where the number of outcomes is different between the parties.}\label{bellbohm}
\end{figure}

The kind of experiment we are considering is sketched in Fig.~\ref{bellbohm}. A source sends out two particle to two distant locations. In each location, a user chooses a possible measurement ($A$, $B$) and registers the outcome ($a$, $b$). The procedure is repeated a large number of times. Later, the two users come together, compare their results and derive the \textit{probability distribution}
\ba
P_{AB}(a,b)\,.\label{pabab}
\ea
This probability distribution is often written as a conditional probability $P(a,b|A,B)$. Ultimately, it's a choice of notation that matters little; I use the notation (\ref{pabab}) to stress that the origin of the statistics on $(a,b)$ is the quantum randomness we want to query, while the statistics on $(A,B)$ are of a very different nature (just the choice of the users on how often to perform each measurement).

A few crucial remarks:
\begin{itemize}
\item[(i)] There is a conceptual distinction between the \textit{users} and the physicists who have constructed the whole setup. To perform the task, the users do not even have to know what state the source has prepared, what physical systems are sent, which measurements are being performed. For the users, the measurement is just a knob they can freely set at any position; $A$ and $B$ refer to the label of the scale in their instruments. In what follows, the names of Alice and Bob will be always referring to the users.
\item[(ii)] The number of possible outcomes $a$ and $b$ plays an important role, but the labeling of the outcomes does not matter: the probability distribution $P(a,b)$ fully characterizes the process of measuring $A$ and $B$, irrespective of whether the outcomes are real numbers, multiples of $\hbar$, complex numbers, colors... Because of this, we are free to choose the most convenient labeling without loss of generality. For instance, if the number of outcomes is two on both sides, then the correlation coefficient is defined as
\ba
E_{AB}&=&P_{AB}(a=b)-P_{AB}(a\neq b)\,.\label{eab}
\ea
This definition is unambiguous and always valid. If in addition, one chooses the labeling $a,b\in\{-1,+1\}$, then one obtains the handy relation $E_{AB}=\moy{ab}_{AB}$ the average of the product of the outcomes.
\end{itemize}

\subsection{Classical mechanisms for correlations}

We are dealing with a \textit{family} of probability distributions: Alice picks up measurement $A$, Bob picks up measurement $B$, and the outcomes are guaranteed to be distributed according to the probability distribution $P_{AB}$. Now, in general, $P_{AB}(a,b)\neq P_A(a)P_B(b)$ where $P_A(a)=\sum_b P_{AB}(a,b)$ and $P_B(b)=\sum_a P_{AB}(a,b)$ are the \textit{marginal distributions}. In other words, random variables distributed according to $P_{AB}$ are correlated. The question that we are going to study is: can these correlations be ascribed to a classical mechanism? The question seems vague, but it becomes less vague once one realizes that there are only \textit{two} classical mechanisms for distributing correlations.

The first and most obvious is \textit{communication}: for instance, the information about Alice's choice of measurement $A$ is available to the particle measured by Bob. This mechanism can be checked by arranging \textit{space-like separation}: loosely speaking, if the two particles reach the measurement devices at the same time and the choice of measurement is done at the very last moment, a signal informing a particle about what is happening in the other location should travel faster than light. If the correlations persist in this configuration, the mechanism of communication becomes highly problematic.

The second mechanism consists in using \textit{pre-established strategies}: each particle might have left the source with a set of instructions, specifying how it should behave in any measurement. Interestingly, this mechanism can explain the behavior of the singlet state in the cases where Alice and Bob choose to perform the same measurement ($P_{MM}$). Indeed, assume that the particles are carrying lists of pre-determined results $\lambda_A=\{a_M\}_M$ and $\lambda_B=\{b_M\}_M$ such that $a_M=-b_M$ for each measurement $M$: one always gets $P_{MM}(a\neq b)=1$, and the local randomness can be easily accounted for by varying the lists at each run. However, it is the milestone result by John Bell in 1964 to prove that the \textit{whole} family of probabilities predicted by quantum physics cannot be reproduced with pre-established strategies \cite{bell64}.

We embark now on a more detailed study of each of the two classical mechanisms and their failure to reproduce the results observed in experiments.

\section{Pre-established strategies (``Local variables'')}

\subsection{The model}\label{secmodel4}

By definition, a pre-established strategy is some hypothetical information $\lambda$ that the particles carry out with themselves from the source. Each particle is supposed to produce its outcome taking into account only this $\lambda$ and the measurement to which it is submitted (plus some possible information encountered along the path, that we neglect for simplicity here). In other words, for given $\lambda$, the two random processes are supposed to be independent: $P_{AB}(a,b|\lambda)=P_{A}(a|\lambda)P_{B}(b|\lambda)$. The only freedom left is the possibility of changing the information $\lambda$ at each run. If $\lambda$ is drawn from a distribution $\rho(\lambda)$, the observed probability distribution will be
\ba
P_{AB}(a,b)&=&\int d\lambda\, \rho(\lambda)\,P_{A}(a|\lambda)P_{B}(b|\lambda)\,.
\ea
Here comes an important mathematical characterization:

\textit{Theorem.} $P_{AB}(a,b)$ can be obtained by pre-established strategies if and only if it can be written as a convex sum of local deterministic strategies.

A \textit{deterministic strategy} is a strategy in which, for each possible measurement, the result is determined. A \textit{local} deterministic strategy is defined by $P_{A}(a|\lambda)=\delta_{a=f(A,\lambda)}$ and $P_{B}(b|\lambda)=\delta_{b=g(B,\lambda)}$. The ``if'' implication is therefore trivial. The ``only if'' implication stems from the fact that any classical random process can be mathematically decomposed as a convex sum of deterministic processes. In other words, for each $\lambda$, there exist an additional random variable $\mu=\mu(\lambda)$ with distribution $\rho'(\mu)$ such that $P_A(a|\lambda)=\int d\mu \rho'(\mu) \delta_{a=f(A,\lambda,\mu)}$. One can therefore just ``enlarge'' the definition of the local variable to $\lambda'=(\lambda,\mu(\lambda))$.

Note that no restriction of the family of pre-established strategies under study is being made. The theorem is a purely mathematical result. It is useful because it allows deriving results by arguing with deterministic strategies, that are very easy to handle; if the result is stable under convex combination, it is automatically guaranteed to hold for \textit{all} possible pre-established strategies, deterministic or not.

\subsection{Two remarks}

Before continuing, it is worth while making two remarks.

\subsubsection{About terminology}

Historically, $\lambda$ was called ``local hidden variable''. While the expression \textit{``local variable''} is ultimately convenient, the adjective ``hidden'' is definitely superfluous and even misleading: quantum physics is at odds with local variables, irrespective of whether they are supposed to be hidden or not. For instance, the local variable may be a description of the total quantum state \cite{gisinreal}. Much more recently, in the interaction between physicists and computer scientists, the name of \textit{``shared randomness''} has also become fashionable to denote pre-established strategies.

Regarding the fact itself, that quantum correlations cannot be attributed to pre-established strategies and local parameters, the linguistic debate is even more involved. Maybe the most precise expression would be ``falsification of crypto-determinism'', as proposed e.g. by Asher Peres \cite{peres}; but it is hardly used. The two most common expressions found in the literature are ``quantum non-locality'' and ``violation of local realism''. Both have their shortcomings, some people in the field have rather strong feeling against one or the other\footnote{For instance, in the language of quantum field theory (which historically pre-dates Bell's inequalities), ``non-locality'' had been used to mean what we call here ``signaling'': if you stick to this meaning, quantum theory and quantum field theory are ``local'', because they are no-signaling.\\ As another example, I tend not to like ``violation of local realism'', because in philosophy ``realism'' is the school of thought that accepts that there is something outside our brain and we can have some convenient knowledge about it. In particular then, a ``philosophical realist'' is someone who accept that ``local realism is violated''.}. You must above all keep in mind that there are plenty of unfortunate expressions in science --- the essence of ``quantum'' physics is the superposition of states, certainly not the discreteness! --- but there is little danger in using them, as long as one knows what their meaning is. I shall use \textit{``non-locality''}, without the ``quantum'', and refer to \textit{``non-local correlations''} as a shortcut for ``probability distributions that cannot be reproduced by pre-established agreement''.

\subsubsection{Pre-established values for single systems}

If quantum systems are considered as a whole, i.e. not consisting of separate sub-systems, one can always find a local variable model that reproduces the correlations. For dimensions larger than two, the Kochen-Specker theorem proves that the local variable model must take into account the whole measurement (it must be ``contextual''); but contextuality is not a problem for local variable theories.

Though its interest may be rather limited, for the sake of illustration, let me introduce a well-known explicit local variable model that reproduces the quantum predictions for one qubit. The goal is to reproduce the statistics of the quantum state $\rho=\demi(\one+\vec{m}\cdot\vec{\sigma})$ under all possible von Neumann measurements, i.e. $P(\pm\vec{a})=\demi(\one\pm\vec{m}\cdot\vec{a})$. The local variable is a vector $\vec{\lambda}=[\sin\theta\cos\phi,\sin\theta\sin\phi,\cos\theta]$ on the surface of the unit sphere; for each realization, $\vec{\lambda}$ is drawn randomly with uniform measure. The rule for the outcome of a measurement is deterministic: the outcome is $r(\vec{a},\vec{\lambda})=\textrm{sign}[(\vec{m}-\vec{\lambda})\cdot\vec{a}]$. Then one can prove that $\moy{r}(\vec{a})=\int_{S^2} r(\vec{a},\vec{\lambda})\,\sin\theta d\theta d\phi\,=\,\vec{m}\cdot\vec{a}$, which is exactly the quantum expectation value (the proof is simple using a geometrical visualization: the integral is the difference between the surface of two spherical hulls).

\subsection{CHSH inequality: derivation}

Let us now present the derivation of the most famous Bell inequality, the one derived by Clauser, Horne, Shimony and Holt and therefore known as CHSH \cite{chsh}. This inequality is defined by the fact that both Alice and Bob can make only two possible measurements (we label them $A,A'$ for Alice, $B,B'$ for Bob) and the outcomes are binary (we choose the labeling $a,b\in\{+1,-1\}$).

Let us first consider a local deterministic strategy $\lambda_D$: here, it is just any list $\lambda_D=(a,a',b,b')$ specifying the two outcomes of Alice and the two outcomes of Bob. If this list is defined, then the number $S(\lambda_D)=(a+a')b+(a-a')b'$ is also defined. By inspection, it is obvious that $S(\lambda_D)$ can only take the values $+2$ or $-2$. If we now take a convex combination of such strategies with distribution $\rho$, it is obvious that $\moy{S}=\int d\lambda_D \rho(\lambda_D)S(\lambda_D)$ must lie between $-2$ and $2$. Moreover, since the average of a sum is the sum of the averages, and since with our labeling $\moy{ab}$ is the correlation coefficient, we have found
\ba
\left|\moy{S}\right|\,=\,\left|E_{AB}+E_{A'B}+E_{AB'}-E_{A'B'}\right|&\leq & 2\,.
\ea
This is the \textit{CHSH inequality}: it holds for all convex combinations of local deterministic strategies and therefore, by virtue of the theorem above, it holds for all pre-established strategies. An important remark here: the derivation of the inequality is particularly simple when the outcomes are labeled $+1$ and $-1$; this is how we did it, and we shall keep this convention in this whole lecture (for the topic of the next lecture, another labeling will prove more convenient). However, the inequality itself is independent of this choice: if one chooses another labeling, the inequality still holds with the general definition (\ref{eab}).

\subsection{CHSH inequality: violation in quantum physics}

In quantum physics, a measurement that can give two outcomes, $+1$ and $-1$, is described by an Hermitian operator with those eigenvalues (possibly degenerate). Therefore
\ba
E_{AB}&\rightarrow& \moy{A\otimes B}
\ea where $A$ and $B$ are two such operators. In other words, in quantum physics $\moy{S}$ becomes the expectation value of the \textit{CHSH operator}
\ba
{\mathcal S}&=&A\otimes B+A'\otimes B+A\otimes B'-A'\otimes B'\,.
\ea
The largest possible value of $\moy{S}$ is therefore\footnote{We are using the well-known fact that the maximal eigenvalue is equal to the largest average value over the set of possible states.} the largest eigenvalue of ${\mathcal S}$. As noted by Tsirelson\footnote{This author used to spell his name as Cirel'son until the mid-eighties.}, in the case of the CHSH inequality it is easy to find a bound for the maximal eigenvalue \cite{tsibound}. Indeed, using the fact that $A^2=A'^2=B^2=B'^2=\one$, one can easily compute ${\mathcal S}^2=4\one\otimes\one +\com{A}{A'}\otimes\com{B}{B'}$. The maximal eigenvalue of $\com{A}{A'}$ cannot exceed 2, because $|\moy{\com{A}{A'}}|\leq |\moy{AA'}|+|\moy{A'A}|$ and the spectrum of both $A$ and $A'$ contains only $+1$ and $-1$. So the maximal eigenvalue of ${\mathcal S}^2$ cannot exceed $8$, which implies that in quantum physics
\ba
\left|\moy{S}\right|&\leq& 2\sqrt{2}\,.
\ea
This bound can be saturated already with two-qubit states. Indeed, consider the correlations of the singlet state $E_{AB}=-\vec{a}\cdot\vec{b}$: by choosing $\vec{a}=\hat{z}$, $\vec{a}\,'=\hat{x}$, $\vec{b}=\frac{1}{\sqrt{2}}(\hat{z}+\hat{x})$, $\vec{b}\,'=\frac{1}{\sqrt{2}}(\hat{z}-\hat{x})$, one obtains $E_{AB}=E_{AB'}=E_{A'B}=-E_{A'B'}=-\frac{1}{\sqrt{2}}$, whence $\left|\moy{S}\right|=2\sqrt{2}$.

Two important remarks:
\begin{itemize}
\item It is remarkable that local variables can be ruled out already in the \textit{simplest} scenario: with the smallest Hilbert space that allows entanglement and with the least possible number of measurements and of outcomes. Indeed, if a party performs only one measurement, one can always find a local variable model that explains the statistics; and a measurement with only one outcome would be obviously trivial.
\item As we have just seen, there is not a unique Bell operator, but a family parametrized by the measurement settings. It is trivial to find ``bad settings'', that don't give any violation even if the state is maximally entangled: for instance, remember that the correlations of the singlet cannot violate any Bell inequality if one requests Alice and Bob to perform the same measurements, i.e. if $A=B$, $A'=B'$. 
\end{itemize}

\subsection{Ask Nature: experiments and loopholes}
\label{loopholes}

Experiments are reviewed in other series of lectures in this school; comprehensive review articles are also available \cite{revsexp,revsexp2}. Here I just address the following question: to which extent alternative models have been falsified.

\begin{figure}[ht]
\includegraphics[scale=0.8]{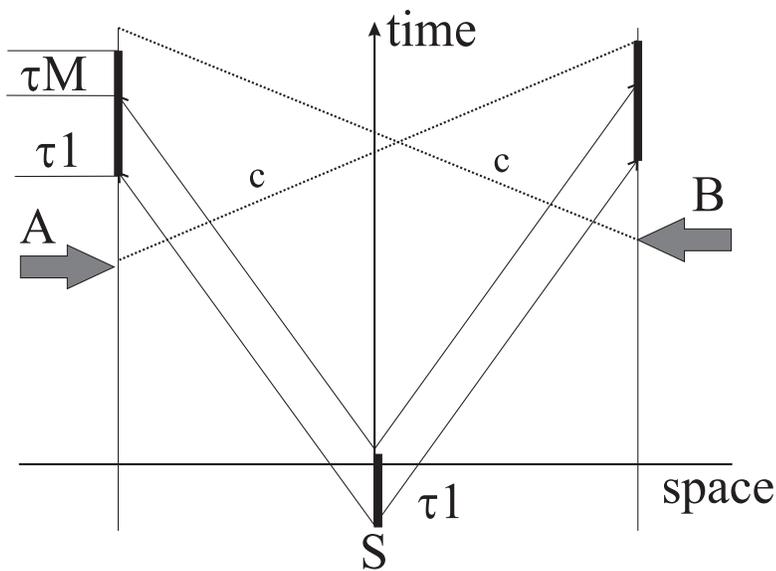}
\caption{Space-time diagram to close the locality loophole: the measurement $A$ must be chosen at a time, such that this information reaches the location of $B$ after the measurement on particle $B$ is completed; and symmetrically. The dotted lines denoted by $c$ represent the light cone in vacuum; the particles, even when they are photons, may propagate at a slower speed (e.g. if they are sent through optical fibers). The emission of a pair happens with some uncertainty $\tau_1$ called single-particle coherence time. The parameter $\tau_M$ indicates the time it takes to perform a measurement, i.e. to ``collapse'' the state: a very problematic notion in this discussion, see text.}\label{locloop}
\end{figure}

The non-locality of quantum correlations is so striking an effect, that it has been scrutinized very closely. Basically two possible \textit{loopholes} have been identified:
\begin{itemize}
\item If one does not arrange the timing properly, the detections may be attributed to a sub-luminal signal: this is called \textit{locality loophole}. In order to close this loophole, the events must be ordered in space-time as in Fig.~\ref{locloop}. Note in particular that the choice of setting on Alice's side must be space-like separated from the end of the measurement on Bob's side, and vice-versa. Now, the ``end of a measurement'' is one of the most fuzzy notions in quantum theory! Consider a photon impinging on a detector: when does quantum coherence leaves place to classical results? Already when the photon generates the first photo-electron? Or when an avalanche of photo-electrons is produced? Or when the result is registered in a computer? There is even an interpretation (Everett's, also called many-worlds) in which no measurement ever happens, the whole evolution of the universe being just a developing of quantum entanglements. All these options are compatible with our current understanding and practice of quantum theory (a fact expressed sometimes by saying that there is a ``quantum measurement problem''). As long as this is the situation, strictly speaking it is impossible to close the locality loophole. However, many physicists adopt the reasonable assumption that the measurement is finished ``not too long time'' after the particle impinges on the detector\footnote{Some theorists having speculated on a link between gravitation and ``collapse'', a recent experiment has bothered connecting the detectors to a piezo-electric device: each time a photon impinges is detected, in addition to the signal being registered by a computer, a \textit{``large'' mass} (a few grams) is set into motion \cite{salart}. If those theoretical speculations have a foundation in nature, this experiment may be extremely meaningful. If not... it had at least the merit of raising the awareness.} and I shall adopt such a position below. In this view, $\tau_M$ is of the order of the microsecond, in which case the distance between Alice and Bob should at least be $300$m.
\item The other possible problem is called \textit{detection loophole}. In all experiments, the violation of Bell's inequalities is measured on the events in which \textit{both} particles have been detected. Since detectors don't have perfect efficiency, there is also a large number of events in which only one particle is detected, while the other has been missed. The detection loophole assumes a form of conspiracy, in which the undetected particle ``chose'' not to be detected after learning to which measurement it was being submitted. In this scenario, if the detection efficiency is not too high, it is pretty simple to produce an apparent violation Bell's inequalities with local variables: when the local variable does not have the desired value, the particle opts not to reply altogether! For CHSH and maximally entangled states, the threshold value of the detection efficiency required to close this loophole is around 80\% (see Tutorial).
\end{itemize} 

At the moment of writing, no experiment has closed both loopholes simultaneously\footnote{The locality loophole can be closed (and has been closed) in experiments with photons; but typical detection efficiencies, including losses, are far from reaching the threshold that would close the detection loophole. In turn, this loophole has been closed in experiments with trapped ions and atoms; but the micrometric separation and long detection times definitely prevent to exclude some sub-luminal communication.}; some groups are aiming at it, but the requirements are really demanding. Is it really worth while? After all, all experimental data are perfectly described by our current physical theory and are in agreement with our understanding of how devices (e.g. detectors) work. By the \textit{usual} standards in physics, the violation of Bell's inequalities by separated entangled pairs is experimentally established beyond doubt. Given the implications of this statement for our world view, though, some people think that \textit{much higher standards} should be applied in this case before making a final claim. I leave it to the readers to choose their camp. I just note here that the detection loophole, so artificial in this context, has recently acquired a much more serious status in a different kind of problems: we shall have to study it in detail in Lecture 6.

\subsection{Entanglement and non-locality}

The literature devoted to Bell's inequalities is huge. All possible generalizations of the CHSH inequality have been presented: more than two measurements, more than two outcomes, more than two particles, and all possible combinations. There are still several open questions \cite{gisquestions,scamet}. Just to give an idea of this complexity, let me review rapidly the development of one of the main questions: \textit{do all entangled states violate a Bell inequality?} This seems an obvious question, but as we shall see, the answers are rather intricate.

The easy part is the following: \textit{all pure entangled states violate some Bell-type inequality}. This statement is known under the name of \textit{Gisin's theorem}: indeed Gisin seems to be the first to have addressed the question, and he proved the statement for bipartite states \cite{gisinthm}; shortly afterwards, Popescu and Rohrlich provided the extension to multipartite states\footnote{The scheme of Popescu and Rohrlich goes as follows: if the system consists of $N$ particles, measurements are performed on $N-2$ of them and the results are communicated to the last two measuring stations: conditioned on this knowledge, the two particles are now in a well-defined bipartite state, so Gisin's original theorem applies. Though it uses communication, the scheme is perfectly valid because the last two parties do \textit{not} communicate with each other: for them, the information communicated by the others acts as pre-established information, which cannot create a violation of Bell's inequalities. For a reason unknown to me, a quite large set of people working in the field are convinced that Gisin's theorem has not been proved in general. Admittedly, for many cases we may not know any compact and elegant inequality that is violated by all pure states; but the scenario of Popescu and Rohrlich does prove the statement in general.} \cite{prgisthm}.

Problems begin when one considers \textit{mixed states}. Even before Gisin proved his theorem, Werner had already shown that there exist mixed states such that (i) they are definitely entangled, in the usual sense that any decomposition as a mixture of pure states must contain some entangled state; (ii) the statistics obtained for all possible von Neumann measurements can be reproduced with pre-established strategies \cite{werner}. Several years later, Barrett showed that the result holds true for the same states even when POVMs are taken into account \cite{barrettpovm}. So \textit{there exist mixed entangled states that do not violate any Bell-type inequality}.

For two qubits, the one-parameter family of Werner states is
\ba
\rho_W&=&W\ket{\Psi^-}\bra{\Psi^-}+(1-W)\frac{\one}{4}
\ea
with $0\leq W\leq 1$. These states are entangled for $W>\frac{1}{3}$. The best extension of Werner's result shows that the statistics of von Neumann measurements are reproducible with local variables for $W\lesssim 0.6595$ \cite{agt}. On the other hand, Werner states provably violate some Bell inequality as soon as $W\gtrsim 0.7056$ \cite{vertesi}. Nobody has been able to close the gap at the moment of writing: this is an open problem, that one may legitimately consider of moderate interest, but that is frustrating in its apparent simplicity\footnote{For the anecdote, let us stress that for many years the known bound for violation was $W>\frac{1}{\sqrt{2}}\sim 0.7071$; this is the bound directly obtained using CHSH. In order to obtain the minor improvement reported above, namely $0.7056$, one needs inequalities that use at least 465 settings per party!}. 

Now, having learned about \textit{distillation of entanglement}, you may legitimately be surprised by Werner's result. Indeed, Werner's states are distillable: why can't one just distill enough entanglement to violate a Bell's inequality? Of course, one can! The previous discussion has been made under the tacit assumption that Alice and Bob do not perform collective measurements, but measure each of their particles individually. This is the way experiments are usually done, but is not the most general scenario allowed by quantum physics. Distillation involves collective measurements, so there is no contradiction between the distillability of Werner's state and Werner's result on local variable models. Actually, it is worth while stressing that the first distillation protocol was invented by Popescu \textit{precisely} in order to extract a violation of Bell's inequalities out of ``local'' Werner states \cite{hiddennl}.

It has recently been shown that \textit{all entangled mixed states (including bound-entangled ones), if submitted to suitable multi-copy processing, can produce statistics that violate some Bell-type inequality} \cite{liang}. This is somehow a comforting result: in this very general scenario, entanglement and the impossibility of pre-established strategies coincide. Still, in my opinion, Werner's result and all its extensions keep their astonishing character: each pair being entangled and produced independently of the others, why does it take complex collective measurements in order to reveal the non-classicality of the source?

\section{Superluminal communication}

\subsection{General considerations}

The issue of superluminal communication as a possible explanation for quantum correlations is delicate. Several physicists think that it is just not worth while addressing in the first place. While I am definitely convinced that no superluminal communication is indeed going on, instead of shunning this topic completely, I prefer to adopt a more pragmatic view based on the following considerations: while local variables have been directly and conclusively disproved as a possible mechanism, superluminal communication seems to be excluded ``only'' on the belief that nothing, really nothing, should propagate faster than light. Now, strictly speaking, relativity forbids faster-than-light propagation of signals that can be used by us (because this would open causality loops and allow signaling in the past). And, as far as I know, nobody has really proved that all possible models based on communication are intrinsically inconsistent --- some serious authors have actually argued the opposite \cite{speedlight,caban}. Therefore, I think it is worth while trying to invent such models and falsify them in experiments.

Still, the family of such models seems to be much more complex than the clear-cut definition of pre-established strategies. Here are a few elements to be taken into account:
\begin{itemize}
\item First of all, when dealing with a hypothetical signal traveling faster than light, one has to decide \textit{in which frame this communication is defined}. There are basically two alternatives: either one considers a \textit{global preferred frame}; or one envisages \textit{separate preferred frames for each particle}. Both alternatives have been explored.
\item By admission of its very founders, Bohmian mechanics can be seen as a theory in which information (the deformation of the quantum potential due to a measurement) propagates at infinite speed in a global preferred frame \cite{bohmhiley}. Now, however problematic the full Bohmian program may be, Bohmian mechanics is mathematically equivalent to quantum mechanics. Therefore, we know that there is at least one model with superluminal communication that cannot be checked against quantum predictions, because its predictions are exactly the same! In other words, there cannot be an analog of Bell's theorem ruling out \textit{all} possible models with communication.
\item When constructing a model, it is customary to enforce the fact that the users should not be able to signal faster than light (i.e. the fact that superluminal communication must be ``hidden''). The models we shall review here are meant to be of this kind; note however that, as soon as one is willing to introduce a preferred frame in physics, there is no compelling reason to enforce no-signaling \cite{eberhard1}.
\end{itemize}

\subsection{Bounds on the speed in a preferred frame}

Let us first suppose that the hypothetical superluminal communication is defined in a \textit{preferred frame}. One has to arrange the detection events to be as simultaneous as possible in this frame; upon observing that the correlations persist, one obtains a lower bound on the speed of this hypothetical communication. Now, we don't know any preferred frame, so which one should one choose? It can be anything, from the local rest frame of the town to the frame in which the cosmic background radiation has no Doppler shift.

It turns out that one does not really have to choose! Indeed, suppose one arranges simultaneity in the rest frame of the town: by virtue of Lorentz transformation\footnote{We assume that standard Lorentz transformation applies to the description of classical events, such as detection; we refer to the discussion above for the problem of defining classical events at all.}, simultaneity is automatically guaranteed in all those frames whose relative speed is orthogonal to the direction A-B. If in addition one arranges the line A-B in the East-West orientation, the rotation of the Earth will scan \textit{all} possible frames in twelve hours!

When experimental data are analyzed, the bounds are striking: the hypothetical communication should travel at speeds that exceed $10000c$ \cite{prefframe,salart2}! This is a very strong suggestion that the mechanism of communication in a preferred frame is not a valid explanation for quantum correlations. 

\subsection{Before-before arrangement in the local frames}

It seems that the previous discussion settles the problem. However, we have another alternative thanks to the inventiveness of Antoine Suarez. This is sometimes known as Suarez-Scarani model, because I happened to help Antoine in formalizing his intuition\footnote{Some years later, I realized that this formalized version was not very brilliant after all: if one tries to extend it to three particles, it leads to signaling \cite{scagisin}! But by then, experiments had already falsified the model anyway \cite{beforebefore}.}. The idea is that there may not be a unique preferred frame: rather, each particle, upon detection, would send out superluminal signals in the rest frame of the measurement device \cite{sua97}.

\begin{figure}[ht]
\includegraphics[scale=0.8]{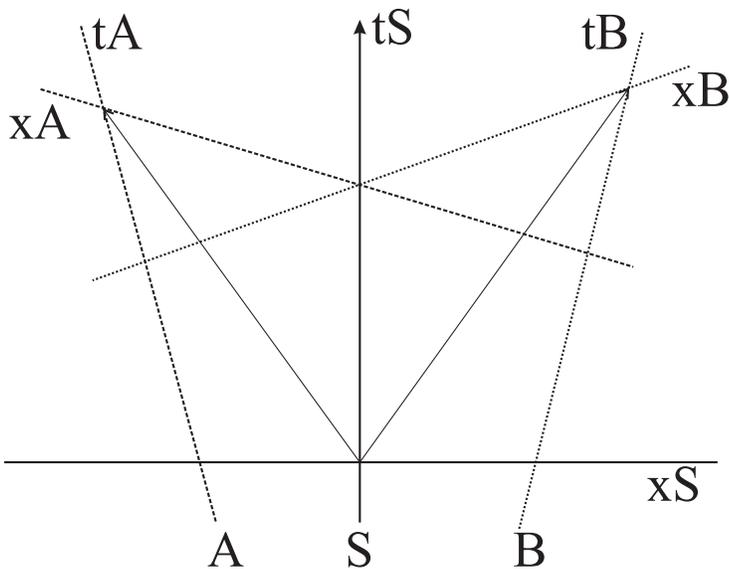}
\caption{Space-time diagram for the before-before experiment (coherence times have been omitted for simplicity, of course they must be taken into account in a detailed analysis). When particle $A$ is detected, in the reference frame of its device particle B had not been detected yet; a symmetrically.}\label{befbef}
\end{figure}

Again, one has to define which is the meaningful measurement device: the choice of the setting, the detector... Whichever one chooses though, how does one falsify such a model? By arranging a \textit{before-before} timing. Consider the space-time diagram sketched in Fig.~\ref{befbef}, representing the situation in which the two measurement devices move away from one another: in that arrangement, each particle arrives before the other one at its respective device! If the model is correct, in this situation one should cease to violate Bell's inequalities. Experiments have been performed and the violation does not disappear \cite{beforebefore,bef2,bef3}: the Suarez-Scarani model adds to the long list of falsified alternative descriptions of quantum phenomena.

\section{Leggett's model}

Let's leave aside the details and admit the failure of the two classical explanations for correlations: something ``non-classical'' is present. Still, one may try to \textit{save as much as possible of a classical world-view}. This is a quite general idea but, as a matter of fact, only one specific model inspired by this line of thought has been proposed so far. This first such proposal is due to Leggett \cite{leggett}, whose initial intuition was rediscovered by the Vienna group \cite{vienna} and has undergone further clarification \cite{cyril}.

Recall that one of the astonishing aspects of entanglement is the fact that a composite system can be in an overall pure state, while none of its components is. One can try to check this statement directly. The problem can be formulated as follows: whether it is possible to find a decomposition
\ba
P_{AB}(a,b)&=&\int d\xi P_{AB}(a,b|\xi)\label{decleggett}
\ea
such that $P_{AB}(a,b|\xi)$ is no-signaling but describes correlations that violate Bell's inequalities (the necessary ``non-classical'' element) but the marginals are like those of a pure, single-particle state. If this is possible, one could have sharp properties for both the composite and the individual systems.

Without entering the details of the derivation, a decomposition like (\ref{decleggett}) fails to recover the quantum probabilities. Experiments have been performed, whose results are in excellent agreement with the latter and falsify therefore the alternative model. In addition, one can prove that, to reproduce the correlations of the singlet state, the marginal of \textit{all} the $P_{AB}(a,b|\xi)$ cannot be anything else than completely random.

From Leggett's model we have learned that not only the full classical mechanisms are ruled out, but also attempts at sneaking back some elements of classicality in an otherwise non-classical theory seem destined to fail. I dare say that this is another remarkable proof of the ``structural accuracy'' of quantum theory: not only, as widely known, this theory is most powerful in its numerical predictions: its structure itself has uncovered properties of reality that we could not have imagined otherwise.

\section{Balance}

Let us summarize the extent of the failure of alternative descriptions to quantum physics, inspired by the phenomenon of quantum correlations at a distance:

\textit{Pre-established agreement:} All possible models based on pre-established agreement (local variables) are incompatible with the quantum predictions. All the experiments are in excellent agreement with the latter; there is a universal consensus that the remaining loopholes are technical issues whose closure is only a matter of time and skill.

\textit{Superluminal communication:} No direct check of all models seems to be possible, and at least one of them is known to be compatible with all quantum predictions. However, all the models that have been tested, together with the implausibility of the assumption itself, clearly uphold the view that no communication is involved in the establishment of quantum correlations (in other words, that there is no \textit{time-ordering} between the correlated events).

There is no third classical way: if one considers (as I do) that we have enough evidence to rule out both pre-established agreement and communication, the conclusion is that \textit{no mechanism in space and time} can reproduce quantum correlations. These correlations ``happen'', we can predict them, but they are a fundamental, irreducible phenomenon. We don't have an explanation in terms of more basic phenomena, and, most remarkably, \textit{we know we shall never find one}.

\section{Tutorials}

\subsection{Problems}

\noindent\textbf{Exercise 4.1}

This calculations shows that all pure entangled states of two qubits violate the CHSH inequality \cite{gisin}. Using the Schmidt decomposition, any two-qubit pure state can be written, in a suitable basis, as $\ket{\psi(\theta)}= \cos\theta\ket{00} +\sin\theta\ket{11}$ with $\cos\theta\geq\sin\theta\geq 0$.
\begin{enumerate}
\item Write down the Bell-CHSH operator $\mathcal{S}$ for $A=\sigma_z$, $A'=\sigma_x$, $B=\cos\beta\sigma_z+\sin\beta\sigma_x$ and $B=\cos\beta\sigma_z-\sin\beta\sigma_x$.
\item Compute $S(\theta|\beta)=\sandwich{\psi(\theta)}{\mathcal{S}}{\psi(\theta)}$, then choose the settings of Bob to achieve the optimal violation $S(\theta)=\max_{\beta}S(\theta|\beta)=2\sqrt{1+\sin^2 2\theta}$.
\end{enumerate}
\textit{Note:} it can be proved that the family of settings above is optimal for the family of states under study. Therefore $S(\theta)$ is the maximal violation of CHSH achievable with the state $\ket{\psi(\theta)}$. For CHSH and two qubits, the optimal violation is also known for mixed states \cite{horo}.\\

\noindent\textbf{Exercise 4.21}

In this exercise, we study the Greenberger-Horne-Zeilinger (GHZ) argument for non-locality \cite{ghz,ghzmermin}. Consider the three-qubit state
\ba
\ket{GHZ}&=&\frac{1}{\sqrt{2}}\left(\ket{000}+\ket{111}\right)\,.
\ea
Compute the four expectation values $\moy{\sigma_x\otimes\sigma_x\otimes\sigma_x}$, $\moy{\sigma_x\otimes\sigma_y\otimes\sigma_y}$, $\moy{\sigma_y\otimes\sigma_x\otimes\sigma_y}$ and $\moy{\sigma_y\otimes\sigma_y\otimes\sigma_x}$ on this state. Try to find a local deterministic strategy that would have such correlations and obtain a contradiction.

\subsection{Solutions}

\noindent\textbf{Exercise 4.1}

Inserting the given settings in the definition of the Bell-CHSH operator, one finds
\ba\mathcal{S}&=&2\cos\beta\,\sigma_z\otimes\sigma_z\,+\, 2\sin\beta\,\sigma_x\otimes\sigma_x\,,\ea whence $\sandwich{\psi(\theta)}{\mathcal{S}}{\psi(\theta)}=\cos\beta+\sin\beta\sin 2\theta$. As well-known, $\max_x(a\cos x+b\sin x)=\sqrt{a^2+b^2}$ obtained for $\cos x = \frac{a}{\sqrt{a^2+b^2}}$; therefore
\ba
S(\theta)=2\sqrt{1+\sin^2 2\theta}&\textrm{ for }& \cos\beta=\frac{1}{\sqrt{1+\sin^2 2\theta}}.
\ea
Note that $S(\theta)$ is always larger than 2, apart from the case $\theta=0$ of a product state.\\

\noindent\textbf{Exercise 4.2}

For the GHZ state, one has
\ba
\begin{array}{lcl} \moy{\sigma_x\otimes\sigma_x\otimes\sigma_x}&=&1\\
\moy{\sigma_x\otimes\sigma_y\otimes\sigma_y}&=&-1\\ \moy{\sigma_y\otimes\sigma_x\otimes\sigma_y}&=&-1\\
\moy{\sigma_y\otimes\sigma_y\otimes\sigma_x}&=&-1
\end{array}
\ea i.e. the state is an eigenvector of those operators. Explicitly, the first line means that only four possible triples of results are possible when all particles are measured in the $x$ direction: $(+,+,+)$, $(+,-,-)$, $(-,+,-)$ and $(-,-,+)$; in the other three cases, the other four triples are possible.

In a local deterministic strategy, all the outcomes should be known in advance, so we want to find six numbers $\{a_X,a_Y;b_X,b_Y;c_X,c_Y\}\in\{+1,-1\}^6$ such that
\ba
\begin{array}{lcl} a_Xb_Xc_X&=&1\\
a_Xb_Yc_Y&=&-1\\ a_Yb_Xc_Y&=&-1\\
a_Yb_Yc_X&=&-1
\end{array}\,.
\ea These relations are however obviously impossible to satisfy (notice for instance that the product of the four lines gives $+1=-1$). So there is no deterministic point satisfying the correlations of the GHZ state. Since these are perfect correlations, no additional element of randomness can create them: in other words, no convex combination of deterministic points can satisfy the correlations either.

\chapter{Quantum correlations (II): the mathematics of no-signaling}

\section{The question of Popescu and Rohrlich}

We saw in the first lecture that the present definition of quantum physics is rather a description of its formalism. Still, there is no shortage of ``typically quantum features'': intrinsic randomness, incompatible measurements and uncertainty relations, no-cloning, teleportation, non-local correlations... Can't one find the physical definition of quantum physics among those?

Popsecu and Rohrlich asked this question for \textit{non-locality without signaling} \cite{PR}: does quantum physics define the set of all probability distributions that are possibly non-local but are compatible with the no-signaling condition? The answer is: NO. Consider the following probability distribution for binary input $A,B\in\{0,1\}$ and binary output $a,b\in\{0,1\}$ (compared to the previous lecture, we use again our freedom of choosing the labels in the most convenient way):
\ba
P_{AB}(a\oplus b=AB)=1&,&P_{AB}(a)=P_{AB}(b)=\demi\,.\label{prbox}
\ea
The distribution is obviously no-signaling: the marginal of Alice does not depend on Bob's measurement (nor on Alice's, in this specific case) and the same for Bob. The correlation says that for $(A,B)\in\{(0,0),(0,1),(1,0)\}$ one has $a=b$ (perfect correlation) while for $(A,B)=(1,1)$ one has $a=b\oplus 1$ (perfect anti-correlation). Therefore, if one evaluates the CHSH expression on this distribution, one finds $\moy{S}=4$! In other words, this innocent-looking probability distribution reaches the largest possible value of CHSH, while we have seen that quantum physics cannot go beyond the Tsirelson bound $\moy{S}=2\sqrt{2}$.

The hypothetical resource that would produce the distribution (\ref{prbox}) has been called \textit{PR-box}. After the Poescu-Rohrlich paper, the PR-box remained rather in the shadow for a few years. Interest in it was revived mainly by two studies. Computer scientists were astonished by the result of van Dam, who noticed that this box would make ``communication complexity'' tasks trivial \cite{vandam}. For physicists, the result of Cerf, Gisin, Massar and Popescu is maybe more appealing \cite{cgmp,degorre}: they proved that the correlations of the singlet can be simulated by local variables plus a single use of the PR-box (thus improving on a previous result by Toner and Bacon \cite{tonerbacon}, who showed the same but using one bit of communication as non-local resource). This latter work raised the hope that the PR-box would play the same role of elementary building-block for non-local distributions, as the singlet plays for quantum states. This hope was later shattered: there are multi-partite non-local distributions that cannot be obtained even if the partners, pairwise, share arbitrarily many PR-boxes \cite{failpr}, and even for bipartite states it seems that full simulation will be impossible beyond the two-qubit case \cite{bacciagaluppi}.

However, the fact that initial hopes are shattered is not new in physics (nor in life, for that matters): creativity is not stopped by that, and indeed, several further interesting results have been obtained along the line of thought started by Popescu and Rohrlich. In order to appreciate them, we have to go a step further. Given the gap between $4$ and $2\sqrt{2}$, one can easily guess that the PR-box is not the only no-signaling resource outside quantum physics: there is actually a continuous family of such objects. The next paragraph is devoted to the formal framework in which no-signaling distributions can be studied.

\section{Formal framework to study no-signaling distributions}

\subsection{A playful interlude}

As a warm up, we consider a family of games in which the two players, Alice and Bob, after learning the rules, are sent to two different locations. There, each of them is submitted to some external input and has to react. The game is won if the reactions are in agreement with the rules that have been fixed in advance.

Specifically, consider the following\footnote{In all the following examples, we assume that there is no correlation between the content of a paper and its acceptance or rejection by a referee. This is not a very strong assumption, especially if the referee has some other interests, like here: winning a game.}\\
\textbf{Game 1:} \textit{each of the players receives a paper to referee. The game is won if, whenever the two players receive the same paper, they produce the same answer (i.e., either both accept or both reject it).}

This game is easy to win: Alice and Bob can just agree in advance that, whatever paper they receive, they will accept it. Admittedly, if many runs of the game are played, this strategy may be a bit boring; but more elaborated ones are possible, for instance: they accept in the first run, reject in the second and third, then accept again in the fourth... Also, the strategy in each round may be much more subtle than a simple ``accept all'' or ``reject all'': for instance, in one round they can decide to ``accept when ([first author name starts with A-M] OR [has been submitted to IJQI]) AND [does not come from CQT]; reject otherwise''. This family of strategies are called \textit{pre-established strategies}. Obviously, there are uncountably many pre-established strategies. Remarkably, though, their set can be bounded, as we already know: it is the convex body, whose extremal points are local deterministic strategies.

Let us now change the rules of the game:\\
\textbf{Game 2:} \textit{each of the players receives a paper to referee. The game is won if the two players produce the same answer \textit{only} when they receive the same paper, and produce different answers otherwise.}

If the set of possible inputs consists only of two papers, the game can still be won by pre-established strategy: the players have just to agree on which paper to accept and which to reject. As soon as there are more than two possible inputs, however, pre-established strategies cannot win the game with certainty. For instance, consider the case where there are three possible papers and write down the conditions for winning: $a_1=b_1$, $a_2=b_2$, $a_3=b_3$, $a_1\neq\{b_2,b_3\}$ etc. It's easy to convince oneself that no set of six numbers $\{a_1,a_2,a_3\}\times\{b_1,b_2,b_3\}$ can fulfill all the conditions. Since no local deterministic strategy can win the game, no pre-established strategy can.

In order to win such a game, in the classical world one has to use the other resource: \textit{communication}. In view of what we know about quantum physics, let us stress here that communication is, in a sense, an exaggerated resource for Game 2. Indeed, with communication, one can win all possible games, for instance\\
\textbf{Game 3:} \textit{each of the players receives a paper to referee. The game is won under the condition: Alice accepts her paper if only if the paper received by Bob has been authored in CQT.}

Why does Game 3 seem more extreme than Game 2? Because Game 3 requires Alice to learn \textit{specific information} about the input received by Bob; while in Game 2, the criteria for winning include only \textit{relations} between Alice's and Bob's inputs and outputs. In other words, communication is intrinsically required to win Game 3; while one might hope to win Game 2 without communication --- with no-signaling resources.

\subsection{Local and no-signaling polytopes: a case study}

Turning now to the formalism, I think it more useful, for the purpose of this school, to do a fully developed case study, rather than giving the general formulas, that those who are going to work in the field will easily find in the literature. Therefore, we focus on the simplest non-trivial case, the same as for the CHSH inequality and the PR-box: two partners, each with binary input $A,B\in\{0,1\}$ and binary output $a,b\in\{0,1\}$.

\subsubsection{Probability space}

The first element to be studied is the \textit{dimensionality of the probability space}: how many numbers are required to specify the four no-signaling probability distributions $P_{AB}(a,b)$ completely? In all, there are sixteen probabilities; but since each of the four $P_{AB}$ must be normalized, we can already reduce to twelve. Actually, a more clever reduction to only \textit{eight} parameters can be achieved exploiting the no-signaling condition. Indeed, note first that the three numbers needed to specify $P_{AB}$ can be chosen as being $P_{AB}(a=0)$, $P_{AB}(b=0)$ and $P_{AB}(a=0,b=0)$, since $P_{AB}(a=0,b=1)=P_{AB}(a=0)-P_{AB}(a=0,b=0)$, $P_{AB}(a=1,b=0)=P_{AB}(b=0)-P_{AB}(a=0,b=0)$ and $P_{AB}(a=1,b=1)$ follows by normalization. Furthermore, because of no-signaling, $P_{AB}(a=0)=P_{A}(a=0)$ for all $A$ and $P_{AB}(b=0)=P_B(b=0)$ for all $B$. Therefore, we are left with eight probabilities that can be conveniently arranged as a table:
\ba
\textbf{P}&=&\begin{array}{c|cc} & P_{B=0}(b=0) & P_{B=1}(b=0)\\\hline P_{A=0}(a=0) & P_{00}(0,0) & P_{01}(0,0)\\ P_{A=1}(a=0) & P_{10}(0,0) & P_{11}(0,0) \end{array}\,.
\ea
For instance, the probability distribution of a PR-box (\ref{prbox}) and the one associated to the best measurements on a maximally entangled state read respectively
\ba
\mathbf{P}_{PR}\,=\,\begin{array}{c|cc} & 1/2 & 1/2\\\hline 1/2 & 1/2 & 1/2\\ 1/2 & 1/2 & 0 \end{array} &,& \mathbf{P}_{ME}\,=\,\begin{array}{c|cc} & 1/2 & 1/2\\\hline 1/2 & \frac{1+1/\sqrt{2}}{4} & \frac{1+1/\sqrt{2}}{4}\\ 1/2 & \frac{1+1/\sqrt{2}}{4} & \frac{1-1/\sqrt{2}}{4} \end{array}\,.
\ea
A priori, Bell's inequalities will appear naturally later in the construction; however, since we have already derived the CHSH inequality in a different way, we can legitimately study here how the inequality looks like in this notation. Noting that $E_{AB}=1-2[P_{AB}(0,1)-P_{AB}(1,0)] =4P_{AB}(0,0)-2P_{A}(a=0)-2P_{B}(b=0)+1$, we find \ban\moy{S}=4\left[P_{00}(0,0)+P_{01}(0,0)+P_{10}(0,0)-P_{11}(0,0)-P_{A=0}(a=0)-P_{B=0}(b=0)\right]+2\,.\ean Remembering that the inequality reads $-2\leq\moy{S}\leq 2$, by simply rearranging the terms we find
\ba
-1\leq &P_{00}(0,0)+P_{01}(0,0)+P_{10}(0,0)-P_{11}(0,0)-P_{A=0}(a=0)-P_{B=0}(b=0)&\leq 0
\ea
i.e. the inequality known as Clauser-Horne (CH) --- which is therefore \textit{strictly equivalent} to CHSH, under the assumption of no-signaling. Written as a table, we have
\ba
\mathbf{T}_{CH}=\begin{array}{c|cc} & -1 & 0\\\hline -1 & 1 & 1\\ 0 & 1 & -1 \end{array}\label{chtable}
\ea
and the inequality reads
\ba
-1\leq & \mathbf{T}_{CH}\cdot\mathbf{P} &\leq 0
\ea
where $\cdot$ represent term-by-term multiplication. For instance, $\mathbf{T}_{CH}\cdot\mathbf{P}_{PR}=\demi$ and $\mathbf{T}_{CH}\cdot\mathbf{P}_{ME}=\frac{1}{\sqrt{2}}-\demi$.

\subsubsection{Local deterministic points (vertices of the local polytope)}

The next step consists in identifying all the \textit{local deterministic strategies}. There are only four deterministic functions $a=f(A)$ from one bit to one bit: $f_1(A)=0$, $f_2(A)=1$, $f_3(A)=A$ and $f_4(A)=A\oplus 1$. Therefore, there are sixteen local deterministic strategies $D^{ij}_{AB}(a,b)=\delta_{a=f_i(A)} \delta_{b=f_j(B)}$. Let us write down explicitly:
\ban
\mathbf{D}_{{11}}\,=\,\begin{array}{c|cc} & 1 & 1\\\hline 1 & 1 & 1\\ 1 & 1 & 1 \end{array} \,,\; \mathbf{D}_{{12}}\,=\,\begin{array}{c|cc} & 0 & 0\\\hline 1 & 0 & 0\\ 1 & 0 & 0 \end{array} \,,\;
\mathbf{D}_{{13}}\,=\,\begin{array}{c|cc} & 1 & 0\\\hline 1 & 1 & 0\\ 1 & 1 & 0 \end{array} \,,\; \mathbf{D}_{{14}}\,=\,\begin{array}{c|cc} & 0 & 1\\\hline 1 & 0 & 1\\ 1 & 0 & 1\end{array}\,,\\
\mathbf{D}_{{21}}\,=\,\begin{array}{c|cc} & 1 & 1\\\hline 0 & 0 & 0\\ 0 & 0 & 0 \end{array} \,,\; \mathbf{D}_{{22}}\,=\,\begin{array}{c|cc} & 0 & 0\\\hline 0 & 0 & 0\\ 0 & 0 & 0 \end{array} \,,\;
\mathbf{D}_{{23}}\,=\,\begin{array}{c|cc} & 1 & 0\\\hline 0 & 0 & 0\\ 0 & 0 & 0 \end{array} \,,\; \mathbf{D}_{{24}}\,=\,\begin{array}{c|cc} & 0 & 1\\\hline 0 & 0 & 0\\ 0 & 0 & 0\end{array}\,,\\
\mathbf{D}_{{31}}\,=\,\begin{array}{c|cc} & 1 & 1\\\hline 1 & 1 & 1\\ 0 & 0 & 0 \end{array} \,,\; \mathbf{D}_{{32}}\,=\,\begin{array}{c|cc} & 0 & 0\\\hline 1 & 0 & 0\\ 0 & 0 & 0 \end{array} \,,\;
\mathbf{D}_{{33}}\,=\,\begin{array}{c|cc} & 1 & 0\\\hline 1 & 1 & 0\\ 0 & 0 & 0 \end{array} \,,\; \mathbf{D}_{{34}}\,=\,\begin{array}{c|cc} & 0 & 1\\\hline 1 & 0 & 1\\ 0 & 0 & 0\end{array}\,,\\
\mathbf{D}_{{41}}\,=\,\begin{array}{c|cc} & 1 & 1\\\hline 0 & 0 & 0\\ 1 & 1 & 1 \end{array} \,,\; \mathbf{D}_{{42}}\,=\,\begin{array}{c|cc} & 0 & 0\\\hline 0 & 0 & 0\\ 1 & 0 & 0 \end{array} \,,\;
\mathbf{D}_{{43}}\,=\,\begin{array}{c|cc} & 1 & 0\\\hline 0 & 0 & 0\\ 1 & 1 & 0 \end{array} \,,\; \mathbf{D}_{{44}}\,=\,\begin{array}{c|cc} & 0 & 1\\\hline 0 & 0 & 0\\ 1 & 0 & 1\end{array}\,,\\
\ean
According to the theorem shown in Lecture 4, we know that the set of local distributions (distributions that can be obtained with pre-established strategies) is the convex set whose extremal points are the deterministic strategies. A convex set with a finite number of extremal point is called ``polytope'', therefore we shall call this set \textit{local polytope}. The extremal points of a polytope are called \textit{vertices}.

\subsubsection{Facets of the local polytope: Bell's inequalities}

The vertices of the local polytope define its \textit{facets}, i.e. the planes that bound the set. If a point is below the facet, the corresponding probability distribution can be reproduced with local variables; if a point is above the facet, it cannot. Therefore, facets are the geometric representation of Bell's inequalities! Actually, in addition to Bell's inequalities, there are many \textit{trivial facets} that can never be violated: conditions like $P_{AB}(a,b)=0$ or $P_{AB}(a,b)=1$ obviously define boundaries within which every local distribution must be found... because \textit{any} distribution must be found therein! We forget about these trivial facets in what follows.

In our simple case, the local polytope is a polygone in an 8-dimensional space, so its facets are 7-dimensional planes. One has to identify the sets of eight points that define one of these planes, then write the equation that define each plane. This can be done by brute force, but we just use some intuition and then quote the known result. The CH inequality indeed defines facets: we have $\mathbf{T}_{CH}\cdot\mathbf{D}=0$ for $\mathbf{D}_{11}$, $\mathbf{D}_{13}$, $\mathbf{D}_{22}$, $\mathbf{D}_{24}$, $\mathbf{D}_{31}$, $\mathbf{D}_{34}$, $\mathbf{D}_{42}$, $\mathbf{D}_{43}$; and one can check that these eight points indeed define a plane of dimension 7. Above this facet, the most non-local point is the PR-box (\ref{prbox}).

Also, $\mathbf{T}_{CH}\cdot\mathbf{D}=-1$ for the other eight deterministic points: this facet is ``opposite''to the previous one, with the local polytope between the two. The most non-local point above this facet is also a PR-box, the one defined by the rule $a\oplus b=AB\oplus 1$. Note that this PR-box is obtained from the ``original'' one by trivial local processing: e.g. Alice flips her outcome.

A simple argument of symmetry gives us immediately six other equivalent facets. Indeed, given a Bell inequality, a relabeling of the inputs and/or the outputs provides another Bell inequality. Their number is most easily counted by studying how many different rules for PR-boxes one can find, and these are obviously $a\oplus b=(A\oplus 1)B$, $a\oplus b=A(B\oplus 1)$ and $a\oplus b=(A\oplus 1)(B\oplus 1)$, with the corresponding opposite facets obtained as before by adding 1 on either side.

Now, it can be proved that there are no other Bell's inequalities for this case: only eight versions of CHSH, each with eight extremal points on the corresponding facet\footnote{Since there are exactly eight points on each facet, the CHSH polytope is a generalized \textit{tetrahedron} (in 3-dimensional space, the tetrahedron is the simplex that has exactly 3 points on each facet). Like in a tetrahedron, one must take convex combinations of all the eight extremal points to define a point ``inside'' the facet; any combination of fewer points defines only points on the ``edges''.} and one single PR-box on top. All these facets being equivalent to the others up to trivial relabeling of the inputs and/or the outputs, it is customary to say that \textit{there is only one Bell inequality for the case of two users, each with binary inputs and outputs; namely, CHSH (or CH)}.

\subsubsection{The no-signaling polytope and the quantum set}

\begin{figure}[ht]
\includegraphics[scale=0.6]{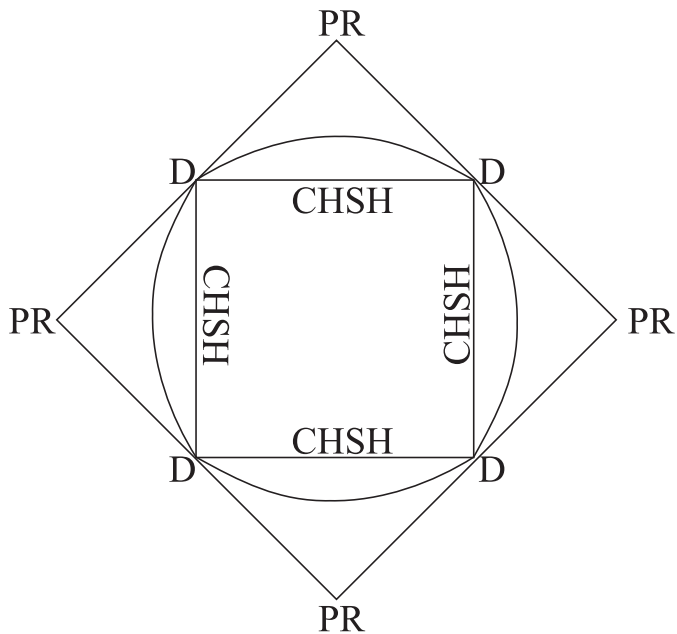}
\caption{Representation of the local polytope, the set of quantum correlations and the no-signaling polytope, for the case under study (2 parties, 2 inputs and 2 outputs per party). The local polytope (inner square) is delimited by versions of the CHSH inequality; the quantum set (round body) exceeds the local polytope, and is contained in the no-signaling polytope (external square), whose extremal points are PR-boxes. Local deterministic points (D) are extremal points of all three sets (not obvious in the drawing, which is just a projection on a two-dimensional slice: the deterministic points do not lie on this slice).}\label{polytopes}
\end{figure}

We have studied at length the set of local distributions. Now we have to say a few words about the two other meaningful sets, namely the no-signaling distributions and the distributions that can be obtained with quantum physics. The image of the probability space is usually drawn as in Fig.~\ref{polytopes}. This drawing is of course a rather poor representation of an 8-dimensional object.

The no-signaling conditions obviously define a convex set (if two distributions are no-signaling, any convex combination will also be no-signaling). It turns out that this set is also a polytope, i.e., it has a finite number of extremal points; obviously, it is called \textit{no-signaling polytope}. All the local deterministic points are also extremal for the no-signaling polytope; but in addition to those, of course, there are some non-local ones, that can in principle be found\footnote{The facets of the no-signaling polytope are rather easy to characterize: they are the ``trivial'' facets that satisfy no-signaling. By intersecting them, a computer program finds the extremal points. So the procedure it's somehow the reverse of the one used to find Bell's inequalities from local deterministic strategies.}. For the simple example we studied, the only additional extremal points are the eight PR-boxes defined above \cite{bletc}.

The \textit{quantum set} is also convex, but is not a polytope: it has an uncountable number of extremal points. Interestingly, at the moment of writing, the shape of this convex body has not been characterized in full generality yet, not even for the simple case under study here. A necessary condition is the following \cite{tsi2,landau,masatsire}: for any probability distribution coming from quantum physics, the correlation coefficients must satisfy an inequality that reminds of CHSH, namely
\ba
\left|\mathrm{arcsin}(E_{AB})+\mathrm{arcsin}(E_{AB'})+\mathrm{arcsin}(E_{A'B})-\mathrm{arcsin}(E_{A'B'})\right|&\leq&\pi\,.
\ea However, this condition is provably not sufficient in general: there are probability distributions that satisfy this inequality but cannot be produced with quantum states \cite{npa}.

\subsubsection{Final remarks}

Let us conclude this section by saying that most of the simple features of this case study are \textit{not} maintained as soon as one considers more complex situations, i.e. more than two inputs, or more than two outputs, or more than two parties.

For a general polytope, the task of finding the facets given the vertices is increasingly complex (in fact, it's provably computationally hard). Many Bell's inequalities have by now been found with computers finding facets --- and possibly sorting the equivalent ones, otherwise the number is just overwhelming\footnote{Note that ``lower'' Bell inequalities remain facets even in more general cases \cite{lifting}: for instance, in any local polytope there are CHSH-like facets that involve only two of the users, two of the inputs and two of the outputs. However, even removing those obvious embeds from lower dimensional polytopes, the number of inequivalent Bell's inequalities grows extremely fast. As an anecdote: Pitowsky and Svozil listed 684 inequalities for the case where the users have each ternary input and binary output \cite{pitowski}. A later inspection by Collins and Gisin \cite{collins} proved that there are actually only \textit{two} non-equivalent ones: CHSH for sure, and a new one that they baptized $I_{3322}$ but that in fact had been already proposed several years earlier in a forgotten paper \cite{froissard}.}. 

Also the characterization of the no-signaling polytope becomes cumbersome. Once moving away from the simplest case, it is no longer the case that all the extremal no-signaling points are equivalent up to symmetries; nor that only one extremal no-signaling point can be found above each facet. The number of objects increases rapidly and the impossibility of visualizing these high-dimensional structures is a powerful deterrent not to embark on such studies unless driven by a very good reason.

\section{The question revisited: why $2\sqrt{2}$?}
\label{2root2}

\subsection{What no-signaling non-local points share with quantum physics}

Most of the features that are usually highlighted as ``typically quantum'' are actually shared by all possible theories that allow non-locality without signaling. This vindicates, in a sense, the intuition of Popescu and Rohrlich: non-locality without signaling is a deep physical principle, to which many observed facts can be attributed.

Here is a rapid list of features that seem to be the share of no-signaling non-local theories, rather than of quantum physics alone:
\begin{itemize}
\item \textit{Intrinsic randomness.} Consider first the PR-box: if one enforces the rule $a\oplus b=xy$ (i.e. maximal algebraic violation of CHSH), it is easy to verify that any prescription for the local statistics other than $P(a=0)=P(b=0)=\demi$ leads to signaling. In other words, the local outcomes of the PR-box \textit{must} be random in order to satisfy no-signaling. Now, this holds in fact for any non-local probability distribution: as soon as $\mathbf{P}$ violates a Bell inequality (be it compatible or not with quantum physics), the local outcomes cannot be deterministic. We'll see in the next Lecture that this can be quantified to provide guaranteed randomness.

\item \textit{No-cloning theorem.} A form of no-cloning theorem can be defined for all non-local no-signaling probability distributions \cite{mas06,bar07}. The formulation goes as follows: suppose there exist $\mathbf{P}_{ABB'}=P(a,b,b'|x,y,y')$ such that $\mathbf{P}_{AB}$ and $\mathbf{P}_{AB'}$ are the same distribution; then $\mathbf{P}_{AB}$ is local. In other words, if $\mathbf{P}_{AB}$ is non-local, one cannot find an extension $\mathbf{P}_{ABB'}$ satisfying no-signaling and such that $\mathbf{P}_{AB'}$ is equal to $\mathbf{P}_{AB}$: the box of B cannot be cloned. The proof is particularly simple in the case of the PR box, see Tutorial.

\item \textit{Possibility of secure cryptography.} It seems that the security of cryptography can be proved only on the basis of no-signaling, without invoking at all the formalism of quantum physics \cite{nskd,agm06}. The exact statement is slightly more involved, since a few technical problems in the security proofs have not been sorted out yet: see the most recent developments for all details \cite{mas08,esther}.

\item \textit{Uncertainty relations, i.e. information-disturbance tradeoff.} Such a tradeoff has been discussed in the context of cryptographic protocols \cite{sca07}.

\item \textit{Teleportation and swapping of correlations.} After a first negative attempt \cite{SHORT-GISIN-POPESCU}, it seems that they can actually be defined as well within the general no-signaling framework \cite{nsswap,paul}. Still work in progress.

\end{itemize}

\subsection{And what they do not.}

Still, there is overwhelming evidence that our world is well described by quantum physics, while there seems to be no evidence of more-than-quantum correlations. Where does the difference lie? This is an open question. Again, I present a rapid list of what is known.

\begin{itemize}

\item \textit{Poor dynamics.} Recall that we are playing with purely kinematical concepts: the $\mathbf{P}$'s are, at least at first sight, the analog of the measurement of an entangled state, not of the state itself! There are suggestions that the allowed dynamics of objects like PR-boxes would be seriously restricted \cite{barrettdyn,barrettdyn2,colbeck}.

\item \textit{Communication tasks becoming trivial.} A few communication tasks have been found, for which quantum physics does not lead to any significant advantage over classical physics, while some more-than-quantum correlations would start helping (the task becoming ultimately trivial if PR-boxes would be available). As mentioned above, the first such example was \textit{communication complexity} \cite{vandam,comcompl,comcompl2}. Two further tasks were proved to become more efficient than allowed in the quantum world as soon as $S>2\sqrt{2}$: \textit{non-local distributed computing} \cite{LPSW} and a form of \textit{random access codes} \cite{infocaus,infocaus2}. These last works formulated the principle of ``information causality'' as a possible criterion to rule more-than-quantum correlations out.

\item \textit{Classical limit.} Finally, it was noticed that most stronger-than-quantum correlations would not recover the classical world in the limit of many copies \cite{nava}. This criterion of ``macroscopic locality'' is related to a family of experiments rather than to an information-theoretical task.

\end{itemize}

\section{Tutorials}

\subsection{Problems}

\noindent\textbf{Exercise 5.1}

Which of the following games can be won by pre-established strategies?
\begin{enumerate}
\item Each of the players receives a paper to referee. The game is won if both players produce always the same answer, unless both papers come from CQT, in which case they must produce different answers.
\item Generalization of Game 2 of the lectures: the two players must produce the same output if and only if they received the same input. \textit{Hint:} Consider the cases where the number of possible inputs is smaller, equal or larger than the number of possible outputs.
\end{enumerate}

\noindent\textbf{Exercise 5.2}

\begin{figure}[ht]
\includegraphics[scale=0.5]{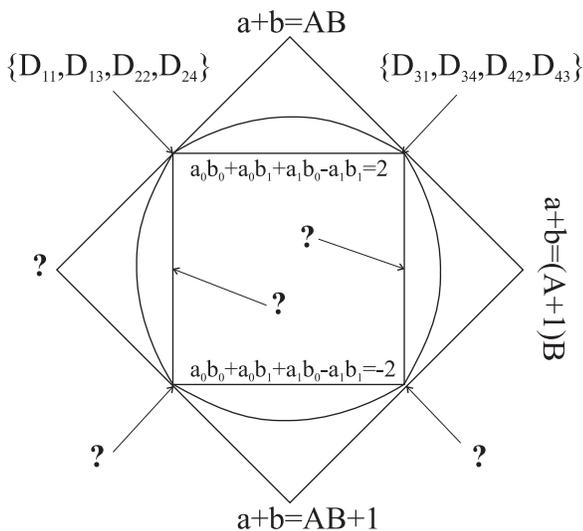}
\caption{Slice of the no-signaling polytope for 2 parties, 2 inputs and 2 outputs per party.}\label{poly2}
\end{figure}

Figure \ref{poly2} represents a two-dimensional slice of the no-signaling polytope for 2 parties, 2 inputs and 2 outputs per party. Add the missing information. \textit{Hint:} in order to assign the deterministic points, re-write the inequality in the form of a table, as done in the text.

\noindent\textbf{Exercise 5.3}

Prove that a tripartite distribution $\mathbf{P}_{ABB'}=P(a,b,b'|x,y,y')$, such that $a\oplus b=xy$ and $a\oplus b'=xy'$, violates the no-signaling constraint. This is the proof of the no-cloning theorem for the specific case of the PR-box.

\subsection{Solutions}

\noindent\textbf{Exercise 5.1}

The first game is obviously non-local: it looks like a generalized PR-box with a large number $N$ of inputs (the possible affiliations) and two outcomes. Interestingly, in the limit $N\rightarrow\infty$, these correlations can be achieved with quantum states \cite{bkp}.

The second game can be won with local strategies if and only if the number of outputs is larger or equal to the number of inputs. Indeed, if this is case, the players agree to output their input, or a function thereof on which they had agreed upon in advance. If this is not the case, a generalization of the argument given in the lectures leads to the conclusion that the game cannot be won.\\

\noindent\textbf{Exercise 5.2}

\begin{figure}[ht]
\includegraphics[scale=0.5]{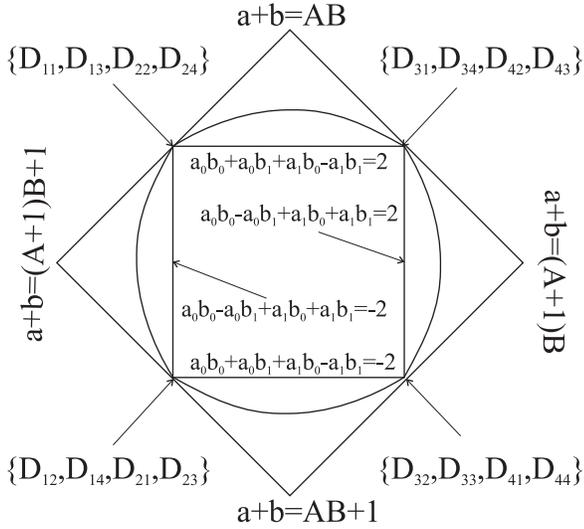}
\caption{Slice of the no-signaling polytope for 2 parties, 2 inputs and 2 outputs per party.}\label{poly3}
\end{figure}

The inequality below the PR-box $a\oplus b=(x+1)y$ is obtained from (\ref{chtable}) by flipping Alice's input, i.e. by flipping the lines:
\ba
\mathbf{T}_{CH}=\begin{array}{c|cc} & -1 & 0\\\hline 0 & 1 & -1\\ -1 & 1 & 1 \end{array}\,.
\ea The rest follows quite immediately; the full result is given in Fig.~\ref{poly3}.\\

\noindent\textbf{Exercise 5.3}

The proof is very simple: the two conditions $a\oplus b=xy$ and $a\oplus b'=xy'$ imply $b\oplus b' = x(y\oplus y')$. So, if both $B$ and $B'$ are given to Bob, he knows $b$, $b'$, $y$ and $y'$ and can therefore reconstruct Alice's input $x$. In other words, signaling is possible from $A$ to $(B,B')$.

\chapter{Quantum correlations (III): the power of Bell}

\section{The model, again}

In section \ref{secmodel4}, the ``local variable'' model was presented in its original flavor (though clarified by some more modern terminology than the one that was used in the 1960's to 1980's): namely, a possible alternative description of nature, supposedly along the line Einstein had in mind. Through the check of Bell's inequalities, experiment is called to rule out this particular model. I find it useful to open this last lecture by re-presenting the model under a different light \cite{gisquestions}, inspired by the present discussions in quantum information science.

The scenario is about two players, Alice and Bob, who are asked to reproduce some quantum statistics without actually measuring the state. Specifically, Alice and Bob are brought together and told that the shall have to reproduce the statistics of a two-particle state $\rho$ under a family of measurements $\{A_j\}$ for Alice and $\{B_k\}$ for Bob. Let's stress that Alice and Bob are given a \textit{complete and exact description on paper} of both the state and the measurements: therefore, they know the statistics $P_{A_j,B_k}(a,b)$ they should obtain for all $j$ and $k$. While still together, they are also allowed to agree on a \textit{common strategy} denoted $s$. After this, Alice and Bob are brought far apart from one another, without possibility of communication, and the game starts: run after run, an external party gives Alice a value of $j$ and Bob a value of $k$, upon which Alice and Bob have to produce outcomes $a$ and $b$ that will satisfy the expected statistics.

Now, Alice and Bob can succeed if and only if the set of $P_{A_j,B_k}(a,b)$ does not violate a Bell-type inequality. While this conclusion should not surprise the reader at this stage, it may be useful to stress that Alice knows quite a lot about Bob's part: she knows the whole state they are supposed to share, she knows the set of measurements Bob will be asked to choose from, and she may have shared a strategy with him. The only thing Alice does not know about Bob is, which $k$ he will receive in each specific run. The same holds for Bob's knowledge about Alice's part. In other words, the ``local variable'' can be as large as $\lambda=\left\{``\rho'',``\{A_j\}'', ``\{B_k\}'',s\right\}$.

The idea can be presented in yet another way: \textit{if Alice and Bob observe a violation of Bell's inequalities and the correlations cannot be attributed to a signal (e.g. because the emission of the signal and the choice of the measurement are space-like separated), then Alice and Bob must be measuring entangled states}. In other words, it makes a difference whether Alice and Bob actually share entangled particles and measure them, or whether they try to simulate the same statistics with purely classical means.

\section{Device-independent: quantum information in a black box}

The fact that violation of Bell's inequalities implies entanglement has been recognized very early in quantum information: in particular, this was precisely Ekert's intuition when he re-discovered quantum cryptography \cite{ekert}. The new development, that I personally find very exciting, is the awareness that \textit{the amount of violation of Bell's inequalities makes it possible to derive quantitative statements on the ``quantumness'' of a black box}. Here follow the specific examples that have been studied to date. Since each case refers to a very limited number of papers, I keep this text quite short, hoping to encourage the reader to read the original papers for all details.

\subsection{Device-independent quantum cryptography}

The first task, to which the idea of device-independent assessment has been applied, is quantum cryptography. This research project amounts precisely at making Ekert's intuition quantitative: if the observed value of the CHSH parameter is $S$, how much information could the eavesdropper Eve have got? Here is \textit{cryptography at its most paranoid}: the authorized partners Alice and Bob could have bought even the source and the measurement devices from the eavesdropper, provided they control the choices of the measurements!

I direct the reader to the original references for a thorough discussion of the scenario, the results and the open issues \cite{devindep,devindepexp,devindep2}. See also \cite{blackpaper} for a more personal assessment of the importance of these results for quantum cryptography in general.

\subsection{Device-independent source characterization}
\label{devindeptomo}

Once the idea of device-independent assessment is understood, a rather obvious task is the direct assessment of the quality of a source: the observed violation of a Bell inequality should be related to the amount of entanglement of the produced state. This question can be given a slightly different turn, inspired by the idea of device-testing \cite{Mayers2004,Magniez06}: if I use the source that yields a given violation for a specific task, how probable it is that the result differs from the one I would have obtained with an ideal source? As the reader reminds from section \ref{ssprobadiscr}, this amounts at finding a bound for the trace distance as a function of the violation of a Bell's inequality. The derivation of such bounds seems very complex even in the simplest case, only partial results are known \cite{bardyn}.

\subsection{Guaranteed randomness}

Yet another intuitive statement that can be turned quantitative: if one violates Bell's inequalities, the individual outcomes must have some intrinsic randomness --- because pseudo-randomness is ultimately deterministic and therefore cannot violate Bell's inequalities! Phrased in a different way: if one observes a violation of Bell's inequalities, one can be sure that the underlying process is random.

This can be an extremely exciting development. As of today, in order to certify a random number generator, one has to look in detail into the complicated process and get somehow convinced that something random (or, at the very least, very hard to predict) is happening. On the contrary, Bell's inequalities provide a very straightforward way of certifying randomness, based on the observed statistics only.

At this point, I may disappoint the reader: at the moment of writing, there is no available document on this topic. To my knowledge, the first such study lies deeply buried in the Ph.D. thesis of Roger Colbeck (Cambridge University); for reasons I refrain from commenting here, this study may never be turned into a publication. A much more thorough investigation by Antonio Ac\'{\i}n, Serge Massar and Stefano Pironio has been announced in several conferences and will hopefully be available soon.

\subsection{Dimension witnesses}

Finally, Bell's inequalities can be used to estimate bounds on the dimension of the Hilbert space of the measured particles. More specifically, one can have the following statement: if an observed violation of some inequality exceeds some threshold, then the system that is being measured must be at least of dimension $D$. The derivation of such statements is technically demanding, in particular because one wants to rule out also POVMs: in order, for instance, to exclude qubits, it is not enough to observe that three outcomes are possible!

As it turns out, the CHSH inequality cannot be used as a dimension witness, because one can already obtain all possible violations with two-qubit states. The first results were obtained independently for inequalities with ternary outcomes \cite{dimwitn} and for inequalities with binary outcomes and more than two measurements \cite{vertesi2}. Note that, in itself, the problem of bounding the dimension of the Hilbert space does not require the use of Bell's inequalities \cite{wehner}.

\section{Detection loophole: a warning}

In all the previous paragraph, I have written ``violation of Bell's inequality'' several times. This expression refers to a \textit{conclusive} violation, i.e. one that is not obtained by post-selection; in other words, \textit{the detection loophole must be closed}. We have already encountered this loophole in Section \ref{loopholes}. There, in the context of experiments, it looked as a very minor point, an absurd conspiracy theory invented by the die-hard of classical mechanisms. Indeed, the vast majority of physicists is certain that the detection loophole will be closed one day: as John Bell himself used to note, it's hard to believe that quantum physics will suddenly be falsified just by increasing the efficiency of our detectors.

However, in device-independent assessment, the situation is quite different. In this scenario, we have not been built devices in order to ask questions to a benevolent (or at least, not malevolent) Nature: rather, we are testing devices built by someone else --- and this someone else may be adversarial. Eve in cryptography, the vendor of an allegedly good source or trusted random number generator: all may have very good reasons to try and cheat us! In other words, if we are not careful, they may engineer a purely classical device that exhibits ``violations of Bell's inequalities'' using the detection loophole. This situation has triggered a revival of interest in the detection loophole. For those who want to know more, there is no review paper available, but the latest work \cite{vpbdet} summarizes pretty well the situation.

\section{Challenges ahead}

When I entered the field of quantum information in the year 2000, the general atmosphere was rather cold about Bell's inequalities. Indeed, the general argument one could hear was: ``we know by now that local hidden variables are not there, it's time to study entanglement theory''. Some of us tried to counter such an argument by vague statements like ``Bell's inequalities uncover one of the most astonishing quantum features, they must be useful for something'' --- not very convincing... but true! It took several years and several detours, but at the moment of writing, the role of Bell's inequalities in quantum information has been vindicated: they are the only tool for device-independent quantum information.

One can scarcely imagine quantum information without the notion of ``qubit'' and still, the assessment on the dimensionality of a quantum system requires a very careful characterization. I would not go as far as suggesting that quantum information will ultimately be formulated without qubits; were it only because there are scenarios in which the physical system is well characterized and there is no reason to be paranoid. But the possibility of device-independent assessment is fascinating, useful... and with plenty of challenges ahead for both theorists and experimentalists!

\section{Tutorials}

\subsection{Problems}

\noindent\textbf{Exercise 6.1}

In this exercise, we consider the detection loophole for maximally entangled state and the CHSH inequality; but the construction can be easily generalized to study other situations, e.g. non-maximally entangled states \cite{eber2} or asymmetric detection efficiencies \cite{asymdetloop,asymgeneva}.

Suppose for simplicity that Alice and Bob have a heralded pair source, so that they know when a pair of particles is expected to arrive. They are not allowed to discard any event, so they agree on the following: if Alice's detector fires, she keeps the result of the detection ($+1$ or $-1$); if not, she arbitrarily decides that the result is $+1$. Bob follows the same strategy. Let now $\eta$ be the efficiency of the detectors (supposed identical for all detectors): then the observed average of CHSH will be
\ba
\moy{S}&=&\eta^2S_2\,+\,2\eta(1-\eta)S_1\,+\,(1-\eta^2)S_0
\ea
where $S_2$ is the expected value of CHSH when both Alice's and Bob's detectors have fired, $S_1$ when only either Alice's or Bob's have fired, and $S_0$ where none has fired.

\begin{enumerate}
\item Supposing everything else is perfect (ideal measurements, no dark counts etc), what are the values of $S_2$, $S_1$ and $S_0$?
\item Prove that $\moy{S}>2$ can be achieved only if $\eta>\frac{2}{\sqrt{2}+1}\approx 82\%$.
\end{enumerate}

\subsection{Solutions}

\noindent\textbf{Exercise 6.1}

If everything is perfect, when both photons are detected we have $S_2=2\sqrt{2}$. However, $S_1=0$: when one party detects and the other does not, there are no correlations between the results; and since the detection comes from half of a maximally entangled state, that bit is completely random. Finally, $S_0=2$: by pre-established agreement, both Alice and Bob output $+1$, thus achieving the local bound. The threshold for $\eta$ follows immediately.

At the address of experimentalists: the fact that, in case of no detection, the local bound is achieved, allows to get rid of the no-detection events altogether (these events are not even defined for most real sources, that are not heralded).


\end{document}